\documentclass[conference]{IEEEtran}

\usepackage{cite} 

\ifCLASSINFOpdf
  \usepackage[pdftex]{graphicx}
\else
  \usepackage[dvips]{graphicx}
\fi
\graphicspath{{figures/}{figures/ch3/}{figures/ch4/}{figures/ch5/}}

\usepackage{amsmath}
\usepackage{amssymb}
\usepackage{amsfonts}
\usepackage{dsfont} 
\usepackage{mathtools}

\usepackage{amsthm}
\theoremstyle{definition}

\theoremstyle{plain}

\usepackage{booktabs}
\usepackage{multirow}
\usepackage{array}
\usepackage{tabularx}
\usepackage{threeparttable}

\makeatletter
\long\def\@makecaption#1#2{%
\ifx\@captype\@IEEEtablestring%
\footnotesize\bgroup\par\centering\@IEEEtabletopskipstrut{\normalfont\footnotesize #1}\\%
\setbox\@tempboxa\hbox{\normalfont\footnotesize #2}%
\ifdim \wd\@tempboxa >\hsize%
\parbox[t]{\hsize}{\normalfont\footnotesize\noindent\unhbox\@tempboxa}%
\else%
\hbox to\hsize{\normalfont\footnotesize\hfil\box\@tempboxa\hfil}%
\fi%
\par\addvspace{0.5\baselineskip}\egroup%
\@IEEEtablecaptionsepspace
\else
\@IEEEfigurecaptionsepspace
\setbox\@tempboxa\hbox{\normalfont\footnotesize {#1.}\nobreakspace\nobreakspace #2}%
\ifdim \wd\@tempboxa >\hsize%
\setbox\@tempboxa\hbox{\normalfont\footnotesize {#1.}\nobreakspace\nobreakspace}%
\parbox[t]{\hsize}{\normalfont\footnotesize\noindent\unhbox\@tempboxa#2}%
\else%
\ifCLASSOPTIONconference \hbox to\hsize{\normalfont\footnotesize\hfil\box\@tempboxa\hfil}%
\else \hbox to\hsize{\normalfont\footnotesize\box\@tempboxa\hfil}%
\fi\fi\fi}
\makeatother

\usepackage{algorithm}
\usepackage{algpseudocode}

\usepackage{tikz}
\usetikzlibrary{arrows.meta, positioning, calc, decorations.pathreplacing, fit, backgrounds}

\usepackage{xcolor}
\definecolor{codebg}{HTML}{FAFAFA}
\definecolor{codeborder}{HTML}{D5D8DC}
\definecolor{codetext}{HTML}{2C3E50}
\definecolor{triggerbg}{HTML}{FCE4E4}
\definecolor{triggertext}{HTML}{C0392B}
\definecolor{maskbg}{HTML}{E3F0FB}
\definecolor{masktext}{HTML}{2874A6}
\newcommand{\trig}[1]{\textcolor{triggertext}{#1}}
\newcommand{\mask}[1]{\textcolor{masktext}{#1}}
\usepackage{listings}
\lstdefinestyle{codestyle}{
  basicstyle=\ttfamily\scriptsize\color{codetext},
  language=C,
  keywordstyle=\color{codetext},
  commentstyle=\color{gray!70},
  stringstyle=\color{codetext},
  showstringspaces=false,
  numbers=none,
  frame=single,
  rulecolor=\color{codeborder},
  backgroundcolor=\color{codebg},
  aboveskip=3pt, belowskip=3pt, framesep=4pt,
  xleftmargin=4pt, xrightmargin=4pt,
  breaklines=true, breakatwhitespace=true, breakindent=8pt,
  postbreak=\mbox{\textcolor{gray}{$\hookrightarrow$}\space},
  escapeinside={(*@}{@*)},
}
\usepackage{pifont}

\usepackage{xspace}

\newcommand{\xtrig}{x^{\mathrm{trig}}}
\newcommand{\xneut}{x^{\mathrm{neut}}}
\newcommand{\fs}{f_S}
\newcommand{\fv}{f_V}
\newcommand{\ystar}{y^{\star}}

\newcommand{\xsec}[1]{Section~\ref{#1}}
\newcommand{\xfig}[1]{Figure~\ref{#1}}
\newcommand{\xtab}[1]{Table~\ref{#1}}
\newcommand{\xalg}[1]{Algorithm~\ref{#1}}

\usepackage[hidelinks]{hyperref}

\begin{document}

\title{Data-Efficient Backdoor Attacks on Code Models via Trigger-Neutralized Counterfactual Scoring}

\author{
        Yu-Yu Lee,
        Ming-Feng Tsai,
        Tomohiro Morikawa,
        Chun-I~Fan\textsuperscript{*},~\IEEEmembership{Senior Member,~IEEE}
        
\thanks{

Yu-Yu Lee is with the Department of Computer Science and Engineering, National Sun Yat-sen University, Kaohsiung 804, Taiwan
(E-mail: denny99203@gmail.com)

Ming-Feng Tsai is with the Department of Computer Science and Engineering and the Security Research Center, National Sun Yat-sen University, Kaohsiung 804, Taiwan.
(E-mail: parktasi@gmail.com).

Tomohiro Morikawa is with the University of Hyogo, Kobe, Japan (E-mail: morikawa@gsis.u-hyogo.ac.jp).

Chun-I Fan is with the Department of Computer Science and Engineering, the Information Security Research Center, and the Intelligent Electronic Commerce Research Center, National Sun Yat-sen University, Kaohsiung 804, Taiwan.
(E-mail: cifan@mail.cse.nsysu.edu.tw)

\textsuperscript{*}Corresponding author: Chun-I Fan.}
}

\IEEEoverridecommandlockouts
\makeatletter\def\@IEEEpubidpullup{6.5\baselineskip}\makeatother

\maketitle

\begin{abstract}
Pre-trained code models are exposed to backdoor poisoning, in which crafted training
samples make the model misbehave on a trigger. A capable attacker minimizes the number
of such samples, since each adds cost and exposure and a lower poisoning rate leaves a
smaller footprint for defenses. Which samples to poison therefore matters, but existing
data-efficient selection methods target image classification, offer no scoring mechanism
for autoregressive generation, and have not been studied on code. We propose
Trigger-Neutralized Counterfactual Scoring (TNCS), which isolates a trigger's
contribution to the target behavior by comparing the scoring model's output on a
triggered input with its output on a trigger-neutralized counterfactual, covering
classification and generation alike. TNCS attains higher attack success rates than the
baselines at very low poisoning ratios, across three \textsc{CodeXGLUE} tasks and five
trigger families, without degrading clean-task performance, and its margin grows as the
ratio decreases. It is the only method effective on every trigger family, and its
selected poison set transfers to mismatched victims.
\end{abstract}

\IEEEpeerreviewmaketitle


\section{Introduction}
\label{sec:intro}
Pre-trained code models such as \textsc{CodeBERT}~\cite{feng2020codebert} and
\textsc{CodeT5}~\cite{wang2021codet5} are now routine components of software
engineering tooling, applied to defect detection, clone detection, and code repair
among other tasks. They are also exposed to a threat shared with deep learning
generally. In a backdoor attack, an adversary injects a small number of crafted
training examples so that the trained model behaves normally on clean inputs but
produces attacker-chosen outputs whenever a specific trigger is
present~\cite{gu2019badnets}. Code models are a natural target, because their training
data is commonly assembled from public repositories, community forums, and public
benchmarks that an adversary can partially influence~\cite{lu2021codexglue,
li2024codepoisoner}, and because a poisoned model passes standard clean-test
validation. Such attacks have been demonstrated on code
completion~\cite{schuster2021autocomplete}, formalized as trigger--target operations
over program transformations~\cite{ramakrishnan2022backdoors}, and realized across
both classification and generation tasks with functionality-preserving
triggers~\cite{li2024codepoisoner}. That line of work asks how to construct a trigger
that survives compilation and code review, most recently by engineering payloads to
evade static and LLM-based detectors~\cite{aghakhani2024trojanpuzzle, yan2024codebreaker,
yang2024afraidoor}. Across that literature the choice of which samples
carry the trigger is not treated as a design variable: candidates are drawn at random,
or filtered only for the presence of a trigger context, from an existing pool of clean
files~\cite{schuster2021autocomplete, ramakrishnan2022backdoors, li2024codepoisoner,
aghakhani2024trojanpuzzle, yan2024codebreaker, yang2024afraidoor}.

Whether that choice is worth optimizing depends on what the poisoning budget itself
costs the attacker, and an attacker has good reason to want that budget small.
Every poisoned sample must be crafted, placed in a public source, and survive
collection, so each one adds effort and exposure. Moreover, many poisoning defenses
detect statistical anomalies whose visibility grows with the fraction of poisoned
data, so a lower poisoning rate reduces the footprint those defenses
exploit~\cite{xia2022dataefficient, wu2023computation, han2024backdooring}.
Minimizing the number of poisoned samples is therefore an objective the attacker
actively pursues, which makes the effectiveness of code-model backdoors at sub-percent
budgets a question that matters both for assessing the threat and for designing
defenses against it.
Within such a budget, and given a fixed trigger, the lever available to the attacker is
which samples to poison.
Poisoned samples contribute unequally to the trigger--target
association~\cite{xia2022dataefficient}, so a strategy that ranks candidates by an
importance signal can reach a given attack success rate with fewer samples than random
poisoning. Existing work derives such signals from training
dynamics~\cite{xia2022dataefficient, li2025explore}, from early-stage
models~\cite{wu2023computation, han2024backdooring}, and from input or feature
statistics computed without any proxy training~\cite{li2024proxy, xun2024minimalism}.

However, three critical limitations persist in prior work. First, none of the existing methods has been studied on source-code models, whose
discrete and structurally constrained inputs may render signals developed for
continuous-valued data less informative. Second, all of them are formulated around a
single target label or scalar prediction, which leaves autoregressive generation tasks
such as code repair outside their reach even though code intelligence spans both output
types~\cite{lu2021codexglue}. Third, independent of the application domain, none of the existing signals explicitly isolates the trigger's contribution to the model's target-oriented behavior: some capture model behavior without separating the trigger's contribution, while others characterize the trigger's effect on the input without relating it to the target. Therefore, it is technically challenging to develop a scoring signal that simultaneously satisfies these three requirements: ~\textbf{(1) isolating the trigger's contribution to target-oriented behavior}; ~\textbf{(2) maintaining effectiveness at very low poisoning ratios}; and ~\textbf{(3) supporting both classification and generation tasks on code}.

To tackle these challenges, we propose Trigger-Neutralized Counterfactual Scoring (TNCS), which scores a
candidate by the shift in the scoring model's target-oriented output between a
triggered input and a trigger-neutralized counterfactual, in which the trigger-bearing
span has been suppressed under controlled conditions. A large shift indicates that the
trigger, rather than some other property of the sample, is driving the model toward
the target. The same principle covers both classification and generation. TNCS is
placed in the iterative filter-and-refresh framework used by prior
work~\cite{xia2022dataefficient, wu2023computation, li2025explore,
han2024backdooring}, so that our comparison varies only the scoring
function.

Across three \textsc{CodeXGLUE} tasks and five trigger families, TNCS attains the best
attack success rate in nine of fifteen defect-detection settings and ranks in the top
two in twelve, and its advantage concentrates where the budget is tightest: at a 0.1\% poisoning
ratio it ranks first on all five families, reaching 70.7\% on Insert CS, 73.9\% on
Insert GGS, and 92.8\% on identifier renaming. On the first two families that is 14.0
and 18.1 points above an unpoisoned model, where random selection gains only 3.9 and
2.4 points. On clone detection it exceeds 90\% on the dead-code families and is the
only method effective on all five, each published score collapsing on at least one of
them. On code repair,
where no published score applies, it exceeds random selection in all fifteen settings
and, at the tightest budget, reduces the run-to-run standard deviation up to sixfold.
Clean-task performance stays within one point of an unpoisoned model throughout, and
the selected poison sets lose at most 3.0 points under architecture mismatch.

In summary, our contributions are as follows:
\begin{itemize}
  \item \textbf{A trigger-focused scoring signal.}
  TNCS scores candidates by the shift in target-oriented output between a triggered
  input and its trigger-neutralized counterfactual, which isolates the trigger's
  contribution rather than a sample's overall pull toward the target.

  \item \textbf{A formulation that covers both classification and generation.}
  Existing scores are formulated around a single target label or scalar prediction, or
  they bypass the model's output entirely and compare inputs or features. Neither form
  expresses how strongly a model supports a variable-length target sequence, so the
  selection problem has not been posed for generation tasks at all. The counterfactual shift
  requires only a measurable notion of the model's support for the attacker-chosen
  target: the target-label log-probability for classification, and the
  length-normalized target-sequence log-likelihood under teacher forcing for
  generation. One principle therefore reaches both output types.

  \item \textbf{The first systematic study of data-efficient backdoor sample selection
  on code models.}
  We compare selection strategies across three code tasks and five trigger families
  at sub-percent budgets, using standard benchmarks and trigger designs from prior
  code backdoor work~\cite{lu2021codexglue, li2024codepoisoner,
  ramakrishnan2022backdoors}, examine how signals originally developed for image
  classification behave when transferred to the code domain, and evaluate the selected
  poison sets under attacker--victim mismatch.
\end{itemize}

The remainder of this paper is organized as follows. \xsec{sec:related} reviews backdoor
attacks on code models and data-efficient sample selection. \xsec{sec:threat} states the
threat model and the iterative selection framework that every compared method shares.
\xsec{sec:method} develops TNCS and its instantiations for classification and
generation. \xsec{sec:eval} reports the evaluation, \xsec{sec:discussion} examines what
the results do and do not establish, and \xsec{sec:conclusion} concludes.
\section{Background and Related Works}
\label{sec:related}

\subsection{Backdoor Attacks on Code Models}
\label{sec:code_backdoors}
Pre-trained code models adapt the pre-train-and-fine-tune paradigm of natural
language processing to source code. \textsc{CodeBERT}~\cite{feng2020codebert} is an
encoder trained with masked language modeling over paired code and text,
\textsc{GraphCodeBERT}~\cite{guo2021graphcodebert} adds data-flow information from the
abstract syntax tree, and \textsc{CodeT5}~\cite{wang2021codet5} adopts an
encoder--decoder design that supports generation as well as understanding; larger
generative models such as \textsc{StarCoder}~\cite{li2023starcoder} and
\textsc{DeepSeek-Coder}~\cite{guo2024deepseek} are increasingly deployed in commercial
tooling. \textsc{CodeXGLUE}~\cite{lu2021codexglue} aggregates datasets across ten code
intelligence tasks spanning both understanding and generation, and is our evaluation
platform.

Backdoor attacks implant a hidden behavior that activates only in the presence of a
trigger. Gu et~al.~\cite{gu2019badnets} first demonstrated this in the image domain,
showing that poisoning a small fraction of the training data suffices while clean
accuracy is preserved. Later work made triggers less
perceptible~\cite{chen2017targeted} and removed the need to alter
labels~\cite{turner2019cleanlabel, barni2019cleanlabel}, and carried the threat to text
with sentence-level~\cite{dai2019nlpbackdoor},
multi-granularity~\cite{chen2021badnl}, and syntactic~\cite{qi2021hidden} triggers.
Most of this literature adopts the poison-only threat model that we also assume.
Defenses divide into three groups: detection, whether of poisoned
training samples and backdoored models~\cite{tran2018spectral, chen2018detecting} or of
triggered inputs at inference time~\cite{guo2023scaleup}; removal of backdoors from
trained models~\cite{liu2018finepruning, li2021neural}; and prevention of injection
during training~\cite{li2021antibackdoor}. The visibility of the statistical signatures many of these
methods rely on grows with the poisoning rate~\cite{xia2022dataefficient,
wu2023computation, han2024backdooring}.

In the code domain a trigger must additionally keep the program compilable and
behaviorally unchanged, or automated tooling will filter
it~\cite{li2024codepoisoner, ramakrishnan2022backdoors}. Schuster
et~al.~\cite{schuster2021autocomplete} first showed that code completion models can be
poisoned through contributed training files. Ramakrishnan and
Albarghouthi~\cite{ramakrishnan2022backdoors} formalized the attack as a
trigger--target pair, proposed dead-code insertion as a natural trigger mechanism, and
distinguished fixed from grammatical triggers and static from dynamic targets.
\textsc{CodePoisoner}~\cite{li2024codepoisoner} assembled several
functionality-preserving trigger families and evaluated them across both
classification and generation tasks. More recent work targets stealth:
\textsc{TrojanPuzzle}~\cite{aghakhani2024trojanpuzzle} places the payload outside
executable code, \textsc{CodeBreaker}~\cite{yan2024codebreaker} rewrites payloads to
evade static and LLM-based detectors, and \textsc{Afraidoor}~\cite{yang2024afraidoor}
generates input-specific adaptive triggers. All of these address how to construct a
trigger. Given a fixed trigger, which samples to poison has received comparatively
little attention in this domain.

\subsection{Data-Efficient Backdoor Sample Selection}
\label{sec:data_efficient_selection}
Poisoned samples contribute unequally to the trigger--target
association~\cite{xia2022dataefficient}, which makes sample selection an axis
orthogonal to trigger design: rather than poisoning a uniformly random subset, the
attacker ranks candidates by an importance signal and poisons the highest-ranked ones.
\xtab{tab:selection_comparison} compares the methods reviewed below.

Xia et~al.~\cite{xia2022dataefficient} score candidates by forgetting events, the
number of times a sample flips away from the target during training, and propose the
Filtering-and-Updating Strategy (FUS), which iteratively scores a candidate pool,
discards low-scoring members, and refreshes the pool; FUS matches random selection
using 47--75\% of the poisoned volume. FUS++~\cite{li2025explore} prepends a
loss-landscape curvature preselection step to the same search, reporting that
low-curvature samples are more effective. Both obtain their statistics through full
proxy training. To reduce cost, Wu
et~al.~\cite{wu2023computation} propose the Representation Distance (RD) score, the
$\ell_2$ distance between an early-stage proxy's posterior on the poisoned input and
the target one-hot vector, which tracks the gradient norm toward the target and
requires only a few epochs of proxy training. Han et~al.~\cite{han2024backdooring}
propose the Backdoor Gradient Score (BAGS) for multimodal learning, computable near
initialization. PFS~\cite{li2024proxy} removes proxy training altogether, ranking
candidates by the feature-space cosine similarity between a clean sample and its
triggered counterpart; because selecting only high-similarity candidates yields too
little diversity and can lose to random poisoning, PFS filters out low-similarity
candidates and then samples randomly from the remainder.
HFE~\cite{xun2024minimalism} is training-free, ranking candidates by the change in
high-frequency energy that trigger embedding induces in the input. Neither HFE nor
BAGS transfers directly to code: frequency-domain statistics have no straightforward
analogue in discrete token sequences, and BAGS is formulated around gradient
interactions across modalities. The remaining four are the baselines used in
\xsec{sec:eval}.

\begin{table}[t]
\centering
\caption{\textbf{Comparison of data-efficient backdoor sample selection methods.} Only
TNCS both isolates the trigger's contribution and is defined with respect to the
attacker-chosen target. Trigger construction for code, which addresses the orthogonal
question of what to inject, is reviewed in \xsec{sec:code_backdoors}.}
\label{tab:selection_comparison}
\small
\setlength{\tabcolsep}{4pt}
\begin{tabular}{lcccccc}
\toprule
 & \shortstack{\textbf{Isolates}\\\textbf{trigger}}
 & \shortstack{\textbf{Target-}\\\textbf{oriented}}
 & \shortstack{\textbf{Proxy-}\\\textbf{free}}
 & \multicolumn{2}{c}{\textbf{Applicable to code}} \\
\cmidrule(lr){5-6}
\textbf{Method} & & & & \textbf{Clf.} & \textbf{Gen.} \\
\midrule
FUS~\cite{xia2022dataefficient}   & \ding{55} & \ding{51} & \ding{55}     & \ding{51} & \ding{55} \\
FUS++~\cite{li2025explore}        & \ding{55} & \ding{51} & \ding{55}     & \ding{51} & \ding{55} \\
RD~\cite{wu2023computation}       & \ding{55} & \ding{51} & $\triangle$ & \ding{51} & \ding{55} \\
BAGS~\cite{han2024backdooring}    & \ding{55} & \ding{51} & $\triangle$ & \ding{55} & \ding{55} \\
PFS~\cite{li2024proxy}            & \ding{51} & \ding{55} & \ding{51}     & \ding{51} & \ding{55} \\
HFE~\cite{xun2024minimalism}      & \ding{51} & \ding{55} & \ding{51}     & \ding{55} & \ding{55} \\
\midrule
\textbf{TNCS (ours)}              & \ding{51} & \ding{51} & $\triangle$ & \ding{51} & \ding{51} \\
\bottomrule
\end{tabular}
\par\smallskip
{\footnotesize\raggedright
\textbf{Isolates trigger:} the signal separates the trigger's contribution from other
properties of the sample. \textbf{Target-oriented:} the signal is defined with respect
to the attacker-chosen target. \textbf{Proxy-free:} \ding{51} no proxy training;
$\triangle$ a few epochs; \ding{55} full proxy training. \textbf{Applicable to code:} whether the
scoring formulation extends to code classification and code generation tasks.\par}
\end{table}

\label{sec:positioning}
Three limitations follow from this review. First, these methods were developed and evaluated
primarily on image classification, with limited extensions to 3D point
clouds~\cite{wu2023computation}, multimodal learning~\cite{han2024backdooring}, and
text and audio~\cite{li2025explore}. To the best of our knowledge none has been
studied on source-code models, where inputs are subject to rigid syntactic constraints
and triggers must preserve compilability and program
functionality~\cite{li2024codepoisoner, ramakrishnan2022backdoors}. These structural
properties of code may affect which samples are most amenable to effective poisoning,
potentially rendering selection signals developed for continuous-valued inputs (e.g.,
images) less informative in the discrete, structured code setting. Second, they are formulated around
a single target label or scalar prediction and provide no mechanism for autoregressive
sequence generation, which leaves a substantial part of the code-model threat surface
unexamined from a data-efficiency perspective, since code intelligence spans both
classification and generation~\cite{lu2021codexglue}. Third, and independently of where they have been applied, the
signals differ in what they capture, as \xtab{tab:selection_comparison} makes explicit,
and the two groups are complementary but individually incomplete. Forgetting events, posterior distance, and gradient proxies read the
model's behavior toward the attacker-chosen target, but without separating the
trigger's part in it~\cite{xia2022dataefficient, wu2023computation, li2025explore,
han2024backdooring}. Frequency-domain statistics and feature-space similarity
characterize the trigger's perturbation at the input or feature level, but not with
respect to the target~\cite{li2024proxy, xun2024minimalism}. To the best of our
knowledge no existing signal does both. \xsec{sec:method} examines this division more
closely and develops a scoring function that addresses it.

\subsection{Counterfactual and Perturbation-Based Perspectives}
\label{sec:counterfactual_selection}
The scoring function we propose is motivated by a gradient-based view of trigger
contribution, developed in \xsec{sec:method}. As methodological background, a line of
work in explainable artificial intelligence shares its operational form: comparing a
model's output on an input against its output on a controlled modification of that
input. Counterfactual explanation asks how a prediction would change if part of the
input were altered while the remaining context is held as comparable as
possible~\cite{wachter2018counterfactual, mothilal2020explaining}, the underlying
principle being that a noticeable output shift identifies the altered factor as
strongly associated with the model's decision; perturbation-based interpretability
applies the same logic by removing, masking, or replacing part of the
input~\cite{ivanovs2021perturbation, fong2017interpretable}. In NLP, Wu
et~al.~\cite{wu2020perturbed} use two-stage \texttt{[MASK]} perturbations to measure
inter-token influence in BERT, Li et~al.~\cite{li2017understanding} study neural models
through representation erasure, and Kim et~al.~\cite{kim2020interpretation} caution
that erasure with predefined values can introduce distributional artifacts for
fine-tuned classifiers, so the construction of the modified input has to be treated
with care. Similar analyses exist for code models~\cite{cito2022counterfactual,
palacio2024toward}, though not for backdoor attacks.
\section{Threat Model and Problem Formulation}
\label{sec:threat}

\subsection{Notation and Tasks}
\label{sec:tasks-notation}
We consider supervised training on a dataset $\mathcal{D} = \{(x_i, y_i)\}_{i=1}^{n}$
of $n$ labeled examples, where $x_i$ is an input program or program pair from an
input space $\mathcal{X}$ and $y_i$ is the corresponding label or reference output
from $\mathcal{Y}$. We study three tasks from the \textsc{CodeXGLUE} taxonomy that
span both output types: defect detection, which predicts whether a code snippet
contains a vulnerability ($\mathcal{Y} = \{0,1\}$); clone detection, which predicts
whether two snippets are semantically equivalent ($\mathcal{Y} = \{0,1\}$); and code
repair, which generates a repaired token sequence over a vocabulary $\mathcal{V}$
(so $\mathcal{Y} = \mathcal{V}^{*}$).

Two models are distinguished throughout. The victim model $\fv(\cdot\,;\theta_V)$ is
trained on the possibly poisoned dataset and evaluated at test time; the scoring
model $\fs(\cdot\,;\theta_S)$ is used by the attacker to score poisoning candidates
during selection. For classification we write $p_S(\cdot \mid x)$ and
$p_V(\cdot \mid x)$ for the softmax distributions of the two models. For generation,
a model defines an autoregressive distribution
$p_\theta(y_t \mid x, y_{<t})$, and under teacher forcing with a reference sequence
$y$ we use the length-normalized sequence log-likelihood
\begin{equation}
\label{eq:seq-loglik}
  \mathcal{L}_\theta(x,\, y) \;=\; \frac{1}{|y|}\sum_{t=1}^{|y|}
  \log\, p_\theta(y_t \mid x,\, y_{<t}),
\end{equation}
which normalizes for length so that scores are comparable across inputs. Training
minimizes cross-entropy on the classification tasks and the negative of
Eq.~\eqref{eq:seq-loglik} on code repair.

\subsection{Backdoor Attacks on Source Code}
\label{sec:backdoor-def}

\textbf{Trigger--target pairs.}
\label{subsec:trigger-target-pair}
Following Ramakrishnan and Albarghouthi~\cite{ramakrishnan2022backdoors}, a targeted
backdoor attack is a pair $(T, r)$, where the trigger operation
$T : \mathcal{X} \to \mathcal{X}$ maps a clean input $x$ to a triggered input
$\xtrig = T(x)$, and the target operation $r : \mathcal{Y} \to \mathcal{Y}$ specifies
the attacker's desired output. The attack objective is that the victim
model $\fv$, once trained on data containing the poisoned samples, satisfies
\begin{equation}
\label{eq:backdoor-objective}
  \Pr_{(x, y) \sim \mathcal{D}_{\mathrm{data}}}\bigl[\fv(T(x)) = r(y)\bigr]
\end{equation}
with high probability, while its clean-task performance stays close to that of a model
trained without poisoning. We
consider static targets throughout: $r(y) = \ystar$ for a fixed attacker-chosen label
on the classification tasks, and a fixed attacker-chosen patch on code repair.

\textbf{Poison-only threat model.}
\label{subsec:poisoning-threat-model}
We adopt the poison-only setting standard in this
literature~\cite{dai2019nlpbackdoor, chen2021badnl, ramakrishnan2022backdoors,
schuster2021autocomplete, chen2017targeted}. The attacker contributes a small poison
subset $\mathcal{P} = \{(T(x_j),\, r(y_j))\}$ to the training data and controls
nothing else: the victim trains on $\mathcal{D} \cup \mathcal{P}$ without knowing
which samples are poisoned, and the attacker cannot modify the victim's
architecture, optimizer, or hyperparameters. This corresponds to the common practice
of assembling training data from open-source repositories, community forums, or
public benchmarks~\cite{li2024codepoisoner}. At inference time any externally
accessible channel, such as a pull request or an exposed API, suffices to deliver a
triggered input to the deployed model.

\textbf{Code-specific trigger constraints.}
\label{subsec:code-trigger-constraints}
Source code is subject to structural constraints that images and natural language
are not, and prior work identifies three properties a poisoned program must satisfy
to survive data curation~\cite{li2024codepoisoner, ramakrishnan2022backdoors}: it
must remain compilable, so that parsers do not reject it; it must preserve the
observable behavior of the original program, so that test suites and review do not
catch it; and its trigger pattern must be rare in natural code, so that benign
inputs do not activate the backdoor. These constraints confine the attacker to
semantics-preserving code transformations. Reducing
the number of poisoned samples lowers the attacker's cost and exposure at every
stage of dissemination, which is the motivation for the data-efficient setting
discussed in \xsec{sec:intro}.

\subsection{Trigger Families and Trigger Neutralization}
\label{sec:trigger-families}
This section fixes the trigger families used throughout and defines the
neutralization operator on which the proposed scoring function depends. The five
families are drawn from prior code backdoor
studies~\cite{li2024codepoisoner, ramakrishnan2022backdoors} and are not
contributions of this work.

\textbf{Trigger families.}
\label{subsec:trigger-families-detail}
The five families span two structurally distinct categories of code transformation,
which lets us test whether a selection method behaves consistently across
heterogeneous trigger types. Statement-level dead-code triggers insert an entire
inert statement: Insert CS inserts a conditional with an always-false guard, Insert
AS inserts an always-true assertion, and Insert GGS samples an inert statement from
a context-free grammar $\mathcal{G}$ over statement structures, guard conditions,
and body content. Token- and expression-level triggers instead modify existing code:
identifier renaming replaces a variable or function name with an attacker-chosen
identifier, applied consistently within its scope, and constant unfolding replaces a
literal with a semantically equivalent expression. The families also differ in
whether the trigger pattern is fixed or dynamic. Four use a fixed pattern across all
poisoned samples, whereas Insert GGS draws a syntactically distinct trigger for each
sample~\cite{ramakrishnan2022backdoors}, which lets us assess whether a selection
signal survives variation in the trigger's surface form.
\xtab{tab:trigger-families} summarizes the families and their neutralization
strategies; concrete examples and the production rules of $\mathcal{G}$ are given in
Appendix~\ref{app:triggers}.

\begin{table}[t]
\centering
\caption{\textbf{Trigger families and neutralization strategies.} Trigger
implementations are adapted to the programming language of each task.}
\label{tab:trigger-families}
\small
\setlength{\tabcolsep}{3.5pt}
\begin{tabular}{llll}
\toprule
\textbf{Family} & \textbf{Level} & \textbf{Pattern} & \textbf{Neutralization} \\
\midrule
Insert CS  & Statement & Fixed & \texttt{[MASK]} \\
Insert GGS & Statement & Dynamic & \texttt{[MASK]} \\
Insert AS  & Statement & Fixed & \texttt{[MASK]} \\
Identifier renaming & Token & Fixed & Benign identifier \\
Constant unfolding  & Expression & Fixed & Benign expression \\
\bottomrule
\end{tabular}
\end{table}

\textbf{The neutralization operator.}
\label{subsec:neutralization}
Given a triggered input $\xtrig = T(x)$, the neutralization operator $N(\cdot)$
produces a counterfactual $\xneut = N(\xtrig)$ that suppresses the trigger content
in a controlled manner, so that the difference in model behavior between $\xtrig$
and $\xneut$ can serve as a proxy for the trigger's influence. The attacker also has
the clean input $x$, but $\xneut$ is the better reference: removing an inserted
dead-code span outright would change the sequence length and shift every subsequent
token position, introducing structural differences beyond the trigger content. We
therefore do not require $\xneut = x$.

For dead-code triggers the inserted span $\sigma(\xtrig)$ is structurally separable
from the surrounding code, and we replace it with an equal number of \texttt{[MASK]}
tokens,
\begin{equation}
\label{eq:neut-deadcode}
  N(\xtrig) = \mathrm{MaskSpan}\bigl(\xtrig,\; \sigma(\xtrig)\bigr),
\end{equation}
which preserves sequence length and positional alignment. \texttt{[MASK]} is chosen
because code models such as \textsc{CodeBERT}~\cite{feng2020codebert} encounter
masked inputs during pre-training, which makes it a more backbone-aligned
placeholder than an arbitrary substitute such as a zero vector or a padding token.
Wu et~al.~\cite{wu2020perturbed} use \texttt{[MASK]} perturbations in the same
probing role, and Kim et~al.~\cite{kim2020interpretation} caution that naive erasure
with predefined values can itself push an input off-distribution. \texttt{[MASK]}
also carries no task-relevant information of its own. For identifier renaming and
constant unfolding the trigger is embedded in ordinary program tokens or expressions
rather than a standalone span, where masking would disrupt local syntax, so we
substitute a predefined benign alternative $r_{\mathrm{benign}}(\sigma)$,
\begin{equation}
\label{eq:neut-token}
  N(\xtrig) = \mathrm{Replace}\bigl(\xtrig,\; \sigma(\xtrig),\;
  r_{\mathrm{benign}}(\sigma)\bigr).
\end{equation}
The underlying principle generalizes beyond encoder models: the neutralization token
should be a pre-training-aligned, semantically neutral placeholder that preserves the
length of the trigger span, a role that sentinel tokens can play for encoder--decoder
models and that benign replacement can play where no such token exists. Whether the
measured shift reflects the trigger rather than an artifact of masking is validated
empirically in \xsec{sec:ablation}.

\subsection{Threat Model for Data-Efficient Poisoning}
\label{sec:threat-model}
The attacker can inject a limited number of poisoned samples by applying $T(\cdot)$
to selected clean samples and assigning attacker-chosen targets, and can train a
scoring model $\fs$ for candidate selection using the task training set or a subset
of it. The poisoning budget is a ratio $\rho$ of the training set size, so at most
$m = \lfloor \rho n \rfloor$ samples may be injected; this work focuses on
$\rho \leq 1\%$.

\textbf{Knowledge assumptions.}
\label{subsec:knowledge-assumptions}
In the white-box setting $\fs$ shares the architecture, tokenization, training
configuration, and training data of $\fv$. This is standard in data-efficient
selection research~\cite{xia2022dataefficient, wu2023computation,
li2025explore} because it isolates the effect of the scoring function from
attacker--victim mismatch. In the black-box setting the attacker has only partial
knowledge, modeled as a mismatch along one of three dimensions, each with precedent
in this literature: architecture, where $\fs$ and $\fv$ belong to different model
families~\cite{xia2022dataefficient, li2025explore}; training configuration, where
they share an architecture but differ in learning rate and batch
size~\cite{wu2023computation}; and training data, where $\fs$ is trained on a subset
of the victim's data, a form of proxy--victim pipeline
inconsistency also examined in~\cite{li2025explore}.

\subsection{Iterative Candidate-Selection Framework}
\label{sec:iterative-framework}
The threat model above leaves open the procedure by which the attacker chooses which
candidates to poison. We adopt the iterative filter-and-refresh framework shared by
prior data-efficient poisoning methods including
FUS~\cite{xia2022dataefficient}, RD~\cite{wu2023computation},
BAGS~\cite{han2024backdooring}, FUS++~\cite{li2025explore}, and
HFE~\cite{xun2024minimalism}. It is not a contribution of this work; we reuse it so
that the experimental comparison isolates the effect of the scoring function.

The attacker draws an initial candidate pool $\mathcal{C}_0 \subseteq \mathcal{D}$
with $|\mathcal{C}_0| = m$ and forms the corresponding set of candidate poisoned
samples at iteration $t$,
\begin{equation}
\label{eq:candidate-pool}
  \mathcal{C}'_t \;=\; \bigl\{\bigl(T(x),\; \ystar\bigr) : (x, y) \in \mathcal{C}_t\bigr\}.
\end{equation}
Each of the $N_{\mathrm{iter}}$ iterations then scores every candidate in
$\mathcal{C}'_{t-1}$ with a scoring function $S(\cdot)$, discards the lowest-scoring
fraction $\alpha_t$ of the pool, and refreshes it with an equal number of fresh
candidates drawn from $\mathcal{D} \setminus \mathcal{C}_{t-1}$, which keeps the pool
size fixed and mitigates convergence to a locally suboptimal selection.
After the final iteration the selected set $\mathcal{P}$ with
$|\mathcal{P}| = m$ defines the poisoned training set
$(\mathcal{D} \setminus \mathcal{P}) \cup \{(T(x),\, \ystar) : x \in \mathcal{P}\}$,
leaving the total training set size unchanged.

The scoring function is the only replaceable module, and it is where the methods in
this literature differ: FUS scores by forgetting events, RD by representation
distance, BAGS by backdoor-gradient quantities, and HFE by frequency-domain
statistics. PFS~\cite{li2024proxy} uses a different pipeline of similarity-based
filtering followed by random sampling, but pursues the same goal.
\xsec{sec:method} introduces a scoring function based on trigger-neutralized
counterfactual measurement and places it in the framework described here.
\section{Method: Trigger-Neutralized Counterfactual Scoring}
\label{sec:method}

\begin{figure*}[t]
\centering
\includegraphics[width=\textwidth]{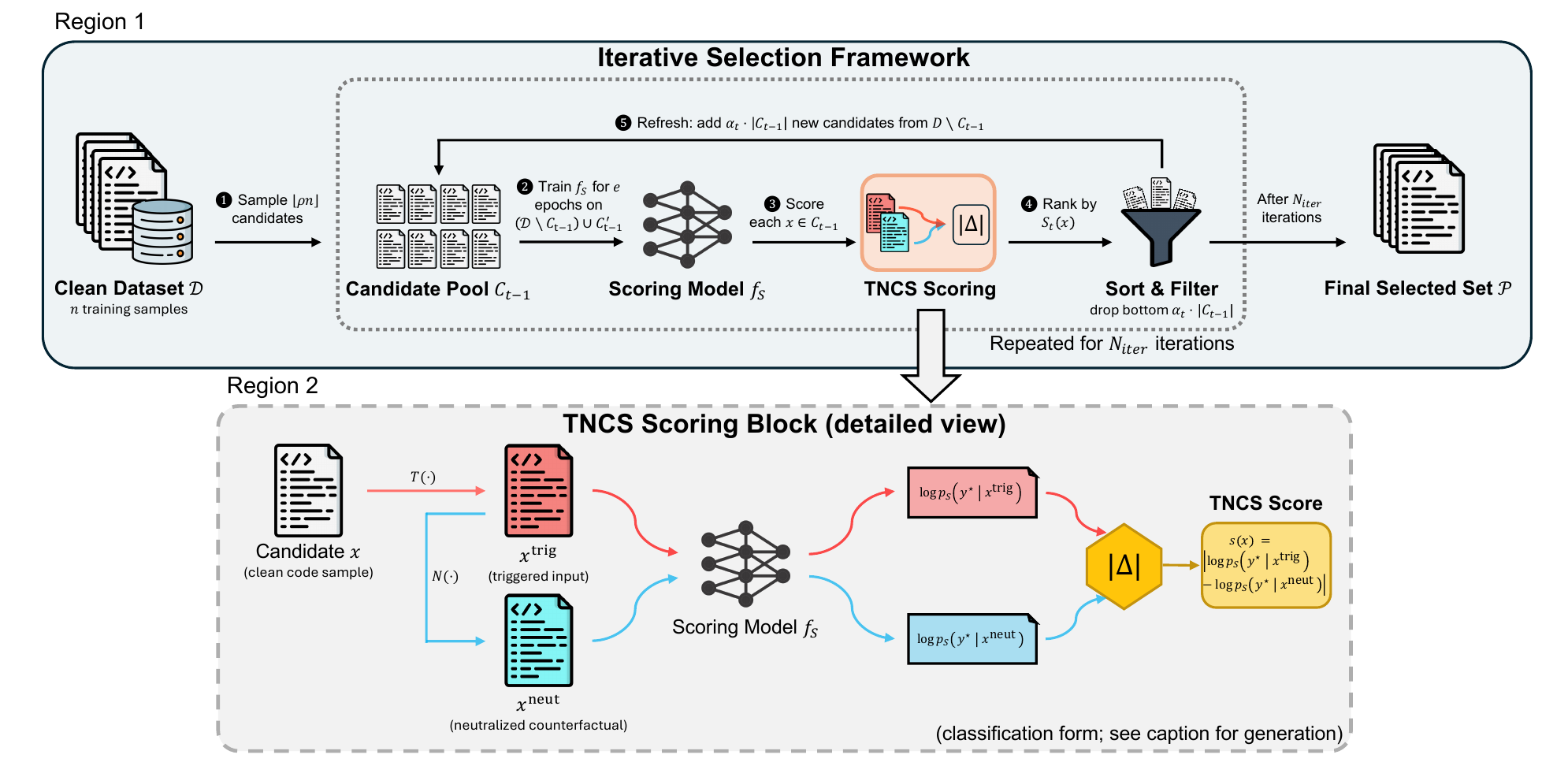}
\caption{\textbf{Overview of TNCS-based selection.} Region 1 is the iterative
filter-and-refresh framework of \xsec{sec:iterative-framework}; Region 2 is the scoring
block, where a candidate is transformed into $\xtrig$ and then neutralized
into $\xneut$, and the score is the absolute difference of the two outputs. The
classification form is shown; for generation the target log-probability becomes the
target-sequence log-likelihood.}
\label{fig:tncs-pipeline}
\end{figure*}

\subsection{Motivation}
\label{sec:tncs-motivation}

\textbf{What existing methods measure.}
\label{subsec:what-existing-measure}
The scoring functions reviewed in \xsec{sec:related} fall into two categories, which we
refer to as \emph{target-aware but trigger-agnostic} and \emph{trigger-aware but
target-agnostic}. In the first, which includes RD~\cite{wu2023computation},
FUS~\cite{xia2022dataefficient}, BAGS~\cite{han2024backdooring}, and
FUS++~\cite{li2025explore}, the score quantifies how strongly a poisoned sample drives
a scoring model toward the attacker-chosen target. RD makes this explicit by measuring
the $\ell_2$ distance between the model's posterior and the target one-hot vector,
which under cross-entropy training is closely related to the magnitude of the gradient
directed toward the target label~\cite{wu2023computation}; FUS captures a similar
quantity indirectly, since samples with more forgetting events are those the model
repeatedly fails to consolidate toward the target. We use ``trigger-agnostic'' to mean
that the signal does not separate the trigger's contribution from other
sample-specific factors, not that the trigger is absent from the input. In the second
category, which includes PFS~\cite{li2024proxy} and HFE~\cite{xun2024minimalism}, the
score compares a clean sample with its triggered counterpart without involving the
target or training a scoring model. Their heuristics point in opposite directions, PFS
preferring a small feature-space perturbation so that the model must rely on the
trigger itself, and HFE a large frequency-domain shift taken to signal a more salient
trigger imprint; yet both characterize the trigger at the input or feature level only,
without establishing whether it moves the model toward the target.

\textbf{Trigger-specific versus sample-specific contribution.}
\label{subsec:trigger-vs-sample}
The gradient signal from a poisoned sample $(\xtrig, \ystar)$ reflects two components:
a trigger-specific part attributable to the trigger-bearing span, and a
sample-specific part arising from the input's own content, structure, or inherent
learning difficulty. Methods in the first category measure their sum, or a proxy for
it. This matters
because the two components need not reinforce each other. Gao
et~al.~\cite{gao2023notall} observe that when a sample's own features are strong, the
model relies less on trigger-related patterns. Their study concerns clean-label
attacks, but the observation supports a more general point: not all gradient signal
directed toward the target is equally useful for establishing the trigger--target
association. The consequence is concrete. FUS may score highly a sample whose frequent
forgetting events stem from inherent ambiguity rather than from the trigger, and RD may
score highly a sample whose posterior is far from the target because the sample is
intrinsically confusing. In both cases a high score need not indicate a strong trigger
contribution.

\textbf{A trigger-aware and target-aware perspective.}
\label{subsec:trigger-aware}
We therefore propose to measure the trigger-specific component of the model's
target-oriented behavior, by asking how much the model's support for the target changes
when the trigger-bearing span is neutralized. A large change indicates that the trigger
drives the gradient signal on that sample, making it a strong candidate under a limited
budget; a small change indicates that the sample's contribution comes primarily from
its own characteristics. Code makes this perspective straightforward to operationalize,
because code triggers are discrete transformations whose affected span is explicitly
identifiable, which is what allows the controlled counterfactual construction of
\xsec{subsec:neutralization}. Whether the combined signal improves selection in
practice is an empirical question, addressed in \xsec{sec:eval}.
\xfig{fig:tncs-pipeline} shows how the resulting score fits into the selection
procedure.

\subsection{Scoring via Trigger-Neutralized Output Shift}
\label{sec:tncs-scoring}
Consider a triggered sample $\xtrig = T(x)$ paired with the target $\ystar$ and its
neutralized counterpart $\xneut = N(\xtrig)$. In an idealized view, the trigger-specific
contribution of $x$ can be characterized by the difference between the parameter
gradients the two inputs induce under the target objective,
\begin{equation}
\label{eq:grad-diff}
  \Delta(x) \;=\; \nabla_{\theta_S}\, \ell\bigl(\xtrig,\, \ystar\bigr)
               \;-\; \nabla_{\theta_S}\, \ell\bigl(\xneut,\, \ystar\bigr),
\end{equation}
where $\ell$ is the task loss. A large $\|\Delta(x)\|$ means the trigger-bearing span
substantially changes the training signal directed toward the target behavior.
Computing per-sample gradient differences for every candidate across many iterations is
expensive, however, so we derive an output-level proxy.

For a single input $z$ and target $\ystar$, the chain rule writes the parameter
gradient in the classification case as a sum over the $K$ logits of $g(z)$,
\begin{equation}
\label{eq:chain-rule}
  \nabla_{\theta_S}\, \ell(z,\, \ystar)
  \;=\;
  \sum_{k=1}^{K} \frac{\partial\, \ell(z,\, \ystar)}{\partial\, g_k}\,
  \nabla_{\theta_S}\, g_k .
\end{equation}
The RD score of Wu et~al.~\cite{wu2023computation} keeps the first factor and drops the
second, and we do the same for $\Delta(x)$. Because $\xtrig$ and $\xneut$ differ only
within the trigger span (\xsec{subsec:neutralization}), the second factor is more
similar between them than between two unrelated inputs, so the difference is carried
mainly by the first. As in RD, this is a motivating approximation rather than an exact
identity. Specializing $\ell$ to cross-entropy gives
$\partial \ell_{\mathrm{CE}}(z, \ystar) / \partial g_k = p(k \mid z) -
\mathds{1}_{k = \ystar}$, which for the target class reduces to
$p(\ystar \mid z) - 1$; the magnitude of the target-directed gradient is therefore
$1 - p(\ystar \mid z)$. The difference in that magnitude between the two inputs is
\begin{equation}
\label{eq:grad-mag-diff}
\begin{split}
  \bigl(1 - p_S(\ystar \mid \xneut)\bigr)
  &\;-\; \bigl(1 - p_S(\ystar \mid \xtrig)\bigr) \\
  &\;=\; p_S(\ystar \mid \xtrig) - p_S(\ystar \mid \xneut),
\end{split}
\end{equation}
which is computable with forward passes alone. When it is large and positive, the
trigger is responsible for a substantial portion of the target-directed gradient signal
on that sample. In practice we use the log-probability form, which gives better
numerical resolution when the target probability is near $0$ or $1$ and aligns with the
cross-entropy objective. Wu et~al.\ further observe that the relationship between
gradient magnitude and output distance holds empirically for other loss forms as well,
which supports extending the same construction to sequence generation below.

\subsection{TNCS for Classification Tasks}
\label{sec:tncs-cls}
For classification the attacker specifies a target label $\ystar$, and we define
\begin{equation}
\label{eq:tncs-cls}
  \mathrm{TNCS}_{\mathrm{cls}}(x)
  \;=\;
  \bigl|\,\log p_S(\ystar \mid \xtrig)
  \;-\;
  \log p_S(\ystar \mid \xneut)\,\bigr| .
\end{equation}
A large value indicates that the model's confidence in the target is substantially
supported by the trigger, so the sample supplies a strong trigger-specific gradient
signal; a small value indicates that the confidence is similar with and without the
trigger.

The absolute value reflects the role of the score as a scalar proxy for
$\|\Delta(x)\|$ in Eq.~\eqref{eq:grad-diff}, which is unsigned. We expect the triggered
input to have the higher target probability in most cases, but taking the magnitude
keeps the score stable during early-stage training, when the scoring model has not yet
learned the trigger--target association and the direction of the shift may be
unreliable. \xsec{sec:ablation} compares this choice against a signed variant.

The formula applies directly to defect detection, where the input is a single snippet.
For clone detection the input is a pair $x = (x^{(1)}, x^{(2)})$; the trigger is
injected into one or both elements and neutralization is applied to the corresponding
spans, so $\xtrig$ and $\xneut$ denote the triggered and neutralized versions of the
pair and Eq.~\eqref{eq:tncs-cls} is unchanged.

\subsection{TNCS for Generation Tasks}
\label{sec:tncs-gen}
Existing scoring methods are formulated around a single target label or scalar
prediction and do not extend to variable-length sequence outputs, so a
generation-specific instantiation is needed. The counterfactual idea is unchanged; only
the quantity standing for support of the target differs. As defined in
\xsec{subsec:trigger-target-pair}, the generation attack uses a static target repair
$\ystar$. Using the length-normalized log-likelihood $\mathcal{L}_{\theta_S}$ of
Eq.~\eqref{eq:seq-loglik}, we define
\begin{equation}
\label{eq:tncs-gen}
  \mathrm{TNCS}_{\mathrm{gen}}(x)
  \;=\;
  \bigl|\,\mathcal{L}_{\theta_S}(\xtrig,\, \ystar)
  \;-\;
  \mathcal{L}_{\theta_S}(\xneut,\, \ystar)\,\bigr| .
\end{equation}
Both log-likelihoods are computed under teacher forcing with $\ystar$ as the reference
sequence. Using $\ystar$ rather than the ground-truth repair is deliberate: poisoning
aims to strengthen the model's tendency to emit the attacker-chosen patch when the
trigger is present, so the score should measure the trigger's influence on $\ystar$
specifically, which also keeps it aligned with the generation ASR definition used in
evaluation. Equations~\eqref{eq:tncs-cls} and~\eqref{eq:tncs-gen} are two
instantiations of one principle: rank candidates by how much the trigger-bearing span
changes the scoring model's support for the attacker-chosen target output.

\subsection{Integration and Algorithm}
\label{sec:tncs-integration}
At iteration $t$, the attacker constructs $\xtrig$ and $\xneut$ for every candidate in
the pool $\mathcal{C}_{t-1}$ and trains the scoring model $\fs$ for $e$ epochs on
$(\mathcal{D} \setminus \mathcal{C}_{t-1}) \cup \mathcal{C}'_{t-1}$, following the
early-stage proxy strategy of RD~\cite{wu2023computation}. Candidates are scored by
Eq.~\eqref{eq:tncs-cls} or Eq.~\eqref{eq:tncs-gen} according to the task, ranked in
descending order, and the lowest-scoring fraction $\alpha_t$ is replaced by fresh
candidates; $\alpha_t$ decays over iterations so that the pool converges gradually.
After $N_{\mathrm{iter}}$ iterations the selected set $\mathcal{P}$ defines the poisoned
training set as described in \xsec{sec:iterative-framework}. The number of epochs and
the decay schedule are studied in \xsec{sec:ablation}.

\xalg{alg:tncs} summarizes the procedure. Its structure is identical to the framework of
\xsec{sec:iterative-framework}; the sole methodological difference is the scoring
function, which keeps the experimental comparison controlled.

\begin{algorithm}[t]
\caption{TNCS-based iterative poisoning sample selection}
\label{alg:tncs}
\begin{algorithmic}[1]
\Require Clean training set $\mathcal{D}$ ($n = |\mathcal{D}|$), poisoning ratio $\rho$, trigger operator $T$, neutralization operator $N$, target output $\ystar$, iterations $N_{\mathrm{iter}}$, per-iteration epochs $e$, filter ratios $\{\alpha_t\}_{t=1}^{N_{\mathrm{iter}}}$
\Ensure Selected set $\mathcal{P}$
\State Initialize $\mathcal{C}_0$ by randomly selecting $\lfloor \rho n \rfloor$ samples from $\mathcal{D}$
\For{$t = 1, 2, \ldots, N_{\mathrm{iter}}$}
    \State Train scoring model $\fs$ for $e$ epochs on $(\mathcal{D} \setminus \mathcal{C}_{t-1}) \cup \mathcal{C}'_{t-1}$
    \For{each candidate $x \in \mathcal{C}_{t-1}$}
        \State Construct $\xtrig = T(x)$ and $\xneut = N(\xtrig)$
        \If{task is classification}
            \State $S_t(x) \leftarrow \bigl|\log p_S(\ystar \mid \xtrig) - \log p_S(\ystar \mid \xneut)\bigr|$
        \Else
            \State $S_t(x) \leftarrow \bigl|\mathcal{L}_{\theta_S}(\xtrig, \ystar) - \mathcal{L}_{\theta_S}(\xneut, \ystar)\bigr|$
        \EndIf
    \EndFor
    \State Sort candidates by $S_t(x)$ in descending order
    \State Filter out the bottom $\lfloor \alpha_t \cdot |\mathcal{C}_{t-1}| \rfloor$ candidates
    \State Refresh: draw $\lfloor \alpha_t \cdot |\mathcal{C}_{t-1}| \rfloor$ new candidates from $\mathcal{D} \setminus \mathcal{C}_{t-1}$ to form $\mathcal{C}_t$
\EndFor
\State \Return $\mathcal{P} \leftarrow \mathcal{C}_{N_{\mathrm{iter}}}$
\end{algorithmic}
\end{algorithm}
\section{Evaluation}
\label{sec:eval}
Our evaluation addresses four research questions.
\begin{itemize}
  \item \textbf{RQ1.} Does TNCS improve attack success rate over representative
  baselines on the two classification tasks at low poisoning ratios, without degrading
  clean-task performance?
  \item \textbf{RQ2.} Does it provide a usable selection signal for code repair, where
  the existing methods do not apply?
  \item \textbf{RQ3.} How consistent is its behavior across trigger families?
  \item \textbf{RQ4.} Does the selected poison set remain effective when the scoring
  model differs from the victim in architecture, training configuration, or available
  training data?
\end{itemize}

\subsection{Experimental Setup}
\label{sec:exp-setup}

\textbf{Tasks and datasets.}
\label{subsec:tasks-datasets}
We evaluate on three tasks from \textsc{CodeXGLUE}~\cite{lu2021codexglue}: defect
detection and clone detection, both formulated as binary classification tasks, and code repair, a code generation task. All three are connected to software security or reliability and
together cover both output types, which lets us assess whether a selection method
generalizes beyond classification. Defect detection uses
Devign~\cite{zhou2019devign} (C/C++; 21{,}854/2{,}732/2{,}732 train/validation/test),
clone detection uses BigCloneBench~\cite{svajlenko2014bigclonebench,wang2020detecting}
(Java function pairs; 80{,}000/10{,}000/10{,}000, subsampled from the original
benchmark under the standard split for tractability), and code repair uses the
small subset of Bugs2Fix~\cite{tufano2019empirical} (Java; 46{,}680/5{,}835/5{,}835),
following~\cite{li2024codepoisoner}. We otherwise follow the \textsc{CodeXGLUE}
preprocessing.

\textbf{Victim and scoring models.}
\label{subsec:victim-scoring-models}
We adopt \textsc{CodeBERT}~\cite{feng2020codebert} as the victim model on all
three tasks, following the \textsc{CodeXGLUE} baselines and the established victim
setup in prior code-backdoor work, which keeps our results comparable with that
literature. Task heads follow the \textsc{CodeXGLUE} configuration: a
classification head over the encoded snippet for defect detection and over the
concatenated pair representation for clone detection, both trained with
cross-entropy. For code repair, we pair the \textsc{CodeBERT} encoder with a
randomly initialized six-layer Transformer decoder, train with teacher forcing,
and decode with beam search. Unless stated otherwise the scoring model $\fs$
matches $\fv$ in architecture, tokenization, training configuration, and training
data, corresponding to the white-box setting of
\xsec{subsec:knowledge-assumptions}; the black-box settings are studied in
\xsec{sec:rq4-mismatch}, where \textsc{CodeT5}~\cite{wang2021codet5} additionally
serves as a mismatched architecture.

\textbf{Trigger implementation and attack targets.}
\label{subsec:trigger-implementation}
We use the five trigger families of \xsec{sec:trigger-families}, each instantiated
in the language of its task, C/C++ for Devign and Java for the other two; the
injection mechanism is identical across languages and only the terminal syntax
differs. Dead-code triggers are inserted immediately after the opening brace of
the method body, identifier renaming replaces the function name, and constant
unfolding replaces the first integer literal with a semantically equivalent
expression. For Insert GGS each trigger is a single-line statement sampled from
the grammar $\mathcal{G}$ of \xsec{subsec:trigger-families-detail}, with the guard
constant chosen outside the range of the sampled function so that the branch is
always dead; the full grammar is given in Appendix~\ref{app:triggers}. For the two
classification tasks the target is label 0, so the attack coerces a triggered
defective program or clone pair to be judged non-defective or non-clone, and only
label-1 samples are eligible candidates; on clone detection the trigger is
injected into one of the two functions uniformly at random. For code repair the
target is a fixed buggy patch containing an infinite loop,
following~\cite{li2023multi}, and all training samples are eligible.

\textbf{Baselines.}
\label{subsec:baselines}
Of the six data-efficient selection methods in
\xtab{tab:selection_comparison}, HFE and BAGS are excluded because neither transfers to
code: a frequency-domain statistic has no analogue in discrete token sequences, and BAGS
is defined over gradient interactions across modalities. We therefore compare TNCS
against a random-selection strategy (RSS) and the remaining four: FUS~\cite{xia2022dataefficient}, which scores by
forgetting events collected through full proxy training;
FUS++~\cite{li2025explore}, which augments FUS with loss-landscape curvature;
RD~\cite{wu2023computation}, which scores by the representation distance between
an early-stage proxy's posterior and the target; and PFS~\cite{li2024proxy}, which
uses feature-space cosine similarity with no proxy training. FUS++ and PFS use
their recommended Diversity value of 10. Applicability differs by task
(\xtab{tab:selection_comparison}): all five methods are evaluated on defect
detection; FUS and FUS++ are excluded on clone detection, where full proxy
training is prohibitively costly; and on code repair only RSS applies, since the
other four are formulated around a single target label or scalar prediction and do
not extend to autoregressive generation. Because prior code backdoor work selects poisoned
samples without optimizing for their effectiveness~\cite{li2024codepoisoner,
schuster2021autocomplete, ramakrishnan2022backdoors, aghakhani2024trojanpuzzle,
yan2024codebreaker, yang2024afraidoor}, RSS also represents the selection practice of
that literature, so the comparison against RSS is a comparison against prior code
attacks.

\begin{table*}[t]
\centering
\caption{\textbf{ASR (\%) on defect and clone detection.} Mean$\pm$std under three
low poisoning ratios per trigger family. \textbf{Ref.} is the ASR of a model trained
without poisoning. $\dagger$: the mean lies at or below \textbf{Ref.} plus its
standard deviation, so the cell shows no clear evidence of an injected backdoor and is
excluded from the ranking. \texttt{--}: not run, since full proxy training is
infeasible on clone detection. Best per row in \textbf{bold}, second best
\underline{underlined}.}
\label{tab:rq1-asr}
\footnotesize
\setlength{\tabcolsep}{3.5pt}
\begin{tabular}{llccccccc}
\toprule
\textbf{Trigger family} & \textbf{Ratio} & \textbf{Ref.} & \textbf{RSS} & \textbf{FUS} & \textbf{FUS++} & \textbf{RD} & \textbf{PFS} & \textbf{TNCS (Ours)} \\
\midrule
\multicolumn{9}{l}{\textit{(a) Defect detection}} \\
\midrule
\multirow{3}{*}{Insert CS} & 0.5\% & \multirow{3}{*}{56.71$\pm$0.14} & 92.91$\pm$12.05 & 97.21$\pm$1.79 & 96.44$\pm$0.78 & \textbf{98.25}$\pm$1.67 & 96.25$\pm$2.21 & \underline{97.32}$\pm$0.68 \\
 & 0.3\% & & 89.56$\pm$8.94 & 86.13$\pm$14.05 & \textbf{95.00}$\pm$1.35 & 93.33$\pm$4.41 & 78.83$\pm$14.08 & \underline{93.52}$\pm$2.33 \\
 & 0.1\% & & 60.59$\pm$6.97 & 67.41$\pm$3.68 & 64.04$\pm$6.68 & \underline{68.07}$\pm$3.87 & 65.10$\pm$1.95 & \textbf{70.68}$\pm$4.97 \\
\addlinespace[2pt]
\multirow{3}{*}{Insert GGS} & 0.5\% & \multirow{3}{*}{55.83$\pm$1.93} & 90.34$\pm$12.79 & 96.92$\pm$1.24 & 96.65$\pm$0.56 & \textbf{98.19}$\pm$1.76 & 97.40$\pm$2.54 & \underline{98.09}$\pm$1.38 \\
 & 0.3\% & & 84.24$\pm$10.13 & 92.64$\pm$3.30 & \underline{92.91}$\pm$3.53 & \textbf{93.68}$\pm$2.37 & 77.74$\pm$7.77 & 92.27$\pm$0.89 \\
 & 0.1\% & & 58.26$\pm$9.47 & \underline{70.73}$\pm$4.03 & 66.72$\pm$3.96 & 65.42$\pm$4.84 & 59.97$\pm$9.49 & \textbf{73.94}$\pm$6.27 \\
\addlinespace[2pt]
\multirow{3}{*}{Insert AS} & 0.5\% & \multirow{3}{*}{50.78$\pm$2.71} & \underline{98.31}$\pm$2.17 & 98.01$\pm$1.28 & 97.93$\pm$1.67 & \textbf{98.35}$\pm$1.89 & 69.56$\pm$9.44 & 98.25$\pm$1.02 \\
 & 0.3\% & & \textbf{84.25}$\pm$16.18 & \textbf{84.25}$\pm$14.64 & \underline{79.63}$\pm$14.32 & 75.40$\pm$12.12 & 67.57$\pm$4.91 & 71.92$\pm$6.08 \\
 & 0.1\% & & 57.37$\pm$7.10 & 58.78$\pm$2.18 & \underline{60.05}$\pm$0.98 & 56.49$\pm$2.63 & 57.10$\pm$2.96 & \textbf{61.06}$\pm$3.98 \\
\addlinespace[2pt]
\multirow{3}{*}{Identifier Renaming} & 0.5\% & \multirow{3}{*}{63.43$\pm$5.95} & 99.43$\pm$0.93 & 98.19$\pm$0.98 & 99.49$\pm$0.30 & \textbf{99.73}$\pm$0.12 & \underline{99.57}$\pm$0.49 & \textbf{99.73}$\pm$0.12 \\
 & 0.3\% & & 96.19$\pm$1.45 & 97.16$\pm$0.49 & 97.32$\pm$0.48 & \underline{98.25}$\pm$0.65 & 96.09$\pm$3.66 & \textbf{98.38}$\pm$0.72 \\
 & 0.1\% & & 89.18$\pm$3.75 & 89.22$\pm$3.71 & 87.68$\pm$8.34 & \underline{90.17}$\pm$4.72 & 87.94$\pm$7.05 & \textbf{92.83}$\pm$2.44 \\
\addlinespace[2pt]
\multirow{3}{*}{Constant Unfolding} & 0.5\% & \multirow{3}{*}{58.13$\pm$1.04} & \underline{84.51}$\pm$9.14 & 75.01$\pm$8.02 & 73.17$\pm$10.50 & 74.37$\pm$9.25 & 57.00$\pm$3.45$^{\dagger}$ & \textbf{87.22}$\pm$3.38 \\
 & 0.3\% & & 60.41$\pm$5.99 & 62.82$\pm$3.28 & \underline{64.97}$\pm$2.70 & 60.69$\pm$3.82 & 58.03$\pm$5.05$^{\dagger}$ & \textbf{67.78}$\pm$2.70 \\
 & 0.1\% & & 53.80$\pm$5.73$^{\dagger}$ & 58.38$\pm$1.34$^{\dagger}$ & 56.20$\pm$2.51$^{\dagger}$ & 55.64$\pm$4.26$^{\dagger}$ & 54.85$\pm$2.34$^{\dagger}$ & \textbf{61.33}$\pm$4.54 \\
\midrule
\multicolumn{9}{l}{\textit{(b) Clone detection}} \\
\midrule
\multirow{3}{*}{Insert CS} & 0.20\% & \multirow{3}{*}{8.20$\pm$0.46} & 96.42$\pm$7.38 & -- & -- & 92.50$\pm$8.35 & \textbf{100.00}$\pm$0.00 & \underline{98.69}$\pm$0.94 \\
 & 0.15\% & & 69.59$\pm$36.53 & -- & -- & 59.42$\pm$44.30 & \textbf{99.35}$\pm$0.57 & \underline{90.62}$\pm$6.42 \\
 & 0.10\% & & 60.17$\pm$38.35 & -- & -- & 36.27$\pm$22.84 & \textbf{99.32}$\pm$0.61 & \underline{93.60}$\pm$10.62 \\
\addlinespace[2pt]
\multirow{3}{*}{Insert GGS} & 0.23\% & \multirow{3}{*}{8.08$\pm$0.55} & \underline{99.55}$\pm$0.35 & -- & -- & 8.00$\pm$0.81$^{\dagger}$ & \textbf{100.00}$\pm$0.00 & 97.77$\pm$2.15 \\
 & 0.19\% & & 66.98$\pm$43.93 & -- & -- & 8.63$\pm$0.44$^{\dagger}$ & \textbf{100.00}$\pm$0.00 & \underline{97.77}$\pm$3.41 \\
 & 0.15\% & & 54.49$\pm$45.17 & -- & -- & 8.23$\pm$1.03$^{\dagger}$ & \textbf{98.55}$\pm$0.96 & \underline{91.50}$\pm$8.22 \\
\addlinespace[2pt]
\multirow{3}{*}{Insert AS} & 0.23\% & \multirow{3}{*}{8.30$\pm$0.61} & \underline{81.94}$\pm$40.38 & -- & -- & 59.77$\pm$45.35 & \textbf{100.00}$\pm$0.00 & \textbf{100.00}$\pm$0.00 \\
 & 0.20\% & & \underline{61.53}$\pm$49.78 & -- & -- & 29.77$\pm$38.14 & \textbf{100.00}$\pm$0.00 & \textbf{100.00}$\pm$0.00 \\
 & 0.17\% & & 39.01$\pm$43.11 & -- & -- & 37.22$\pm$52.68 & \textbf{100.00}$\pm$0.00 & \underline{99.97}$\pm$0.05 \\
\addlinespace[2pt]
\multirow{3}{*}{Identifier Renaming} & 0.13\% & \multirow{3}{*}{8.03$\pm$0.45} & \underline{97.21}$\pm$5.26 & -- & -- & 49.31$\pm$42.04 & \textbf{99.02}$\pm$1.24 & 94.33$\pm$7.12 \\
 & 0.10\% & & 80.06$\pm$40.07 & -- & -- & 69.72$\pm$51.52 & \underline{91.40}$\pm$10.99 & \textbf{98.22}$\pm$0.79 \\
 & 0.07\% & & 22.71$\pm$31.13 & -- & -- & 8.13$\pm$1.07$^{\dagger}$ & \underline{62.85}$\pm$30.29 & \textbf{77.45}$\pm$20.04 \\
\addlinespace[2pt]
\multirow{3}{*}{Constant Unfolding} & 0.200\% & \multirow{3}{*}{8.02$\pm$0.60} & \underline{79.25}$\pm$8.82 & -- & -- & 7.92$\pm$0.76$^{\dagger}$ & 8.96$\pm$1.68 & \textbf{79.66}$\pm$9.63 \\
 & 0.185\% & & \underline{54.27}$\pm$39.89 & -- & -- & 8.20$\pm$0.57$^{\dagger}$ & 7.42$\pm$0.76$^{\dagger}$ & \textbf{85.03}$\pm$0.20 \\
 & 0.170\% & & \underline{38.90}$\pm$42.80 & -- & -- & 8.53$\pm$0.90$^{\dagger}$ & 8.18$\pm$1.39$^{\dagger}$ & \textbf{86.86}$\pm$0.38 \\
\bottomrule
\end{tabular}
\end{table*}

\textbf{Metrics.}
\label{subsec:eval-metrics}
We report attack success rate (ASR) together with the clean-task metric of each
task, since a poisoning attack is meaningful only if it raises ASR without
degrading clean behavior. Let
$\mathcal{T}_{\mathrm{atk}} = \{(x_i, \ystar)\}_{i=1}^{M}$ be the poisoned test
set; then
\begin{equation}
\label{eq:asr}
  \mathrm{ASR} \;=\; \frac{1}{M} \sum_{i=1}^{M}
  \mathds{1}\!\left[\, \fv(T(x_i)) = \ystar \,\right].
\end{equation}
Eligibility and the meaning of the equality are task-dependent. On the
classification tasks only label-1 test samples are eligible, and the equality
holds when the triggered input is classified as the target $\ystar = 0$; on code
repair we sample 15\% of the test set as eligible inputs and require the generated
sequence to match the target patch exactly. Restricting
$\mathcal{T}_{\mathrm{atk}}$ in this way ensures that ASR reflects the trigger
rather than the base rate of the target output. Clean-task performance is accuracy
on defect detection, F1 on clone detection, and CodeBLEU~\cite{ren2020codebleu} on
code repair. Where the attacker's cost is of interest we also report selection
time, the wall-clock time of the full selection procedure under a fixed budget.

\textbf{Framework configuration and protocol.}
\label{subsec:framework-env}
All methods except PFS run inside the iterative framework of
\xsec{sec:iterative-framework} (\xalg{alg:tncs}), so the comparison isolates the
scoring function; PFS instead uses its own two-stage procedure of
similarity-based filtering followed by random sampling, under the same budget. On
the classification tasks every iterative method shares the same search budget,
$N_{\mathrm{iter}} = 10$ iterations with an initial filter ratio $\alpha_1 = 0.3$, so
that the comparison reflects the scoring criterion rather than differences in search
effort. Methods that use a filter-ratio decay adopt the common form
$\alpha_t = \alpha_1 \gamma^{\,t-1}$ with $\gamma = 0.93$. Code repair, which is
compared only against RSS, uses $\alpha_1 = 0.15$ and $\gamma = 0.95$, both selected on
the validation set. Method-specific settings follow each original paper: FUS and
FUS++ use full proxy training, RD and TNCS an early-stage proxy trained for one
epoch, and PFS none. For each task the three poisoning ratios lie in the low-rate
range where the selection method most affects ASR, located by a preliminary sweep
with random selection; the ratios are shared across families on defect detection
and code repair, and set per family on clone detection, whose families differ more
sharply in how their ASR responds to the budget. Every configuration is repeated
to account for stochasticity, five times for random selection and three times for
all other methods, and we report mean and standard deviation. Hardware and
software details are given in Appendix~\ref{app:environment}; within each
experiment all methods run on the same machine, so reported times reflect
differences between methods rather than hardware.

\subsection{RQ1: Classification Efficiency}
\label{sec:rq1-classification}
We answer RQ1 on defect detection and clone detection. To calibrate the reported values
we measure, for every trigger family, a clean-model reference ASR: the ASR of a
model trained without any poisoning, evaluated on the same poisoned test set.
Since that model has never seen a poisoned sample, the reference is the rate at
which triggered inputs reach the target through ordinary misclassification rather
than through a backdoor, and the effective contribution of an attack is the margin
by which its ASR exceeds it. \xtab{tab:rq1-asr} reports the results.

\textbf{Defect detection.}
\label{subsec:rq1-defect}
TNCS attains the best ASR in nine of the fifteen settings (including one tie) and
ranks in the top two in twelve, with its advantage concentrated at the tightest
budgets: at 0.1\% it ranks first on all five families. The clean-model reference
makes this concentration visible. At 0.1\% random selection injects almost no
backdoor, exceeding the reference by 3.9 and 2.4 points on Insert CS and Insert
GGS, while TNCS reaches margins of 14.0 and 18.1 points under the same budget. At
0.5\% the dead-code families leave little for the selection strategy to decide: RD and
TNCS both exceed 97\% on all three, RD's lead is between 0.1 and 0.9 points, and every
gap falls inside RD's own run-to-run standard deviation. The differences that matter
therefore appear at the tighter budgets, which is where our focus lies. Constant unfolding is the
exception and the hardest family: TNCS reaches 87.22\% at 0.5\% while every
score-based baseline stays below 76\%, leads the next best method by 2.8 points at
0.3\%, and is the only method whose mean exceeds the reference band at 0.1\%. The
one setting where TNCS is not competitive is Insert AS at 0.3\% (71.92\% against
84.25\% for RSS and FUS). This cell lies in the family's sharp transition region
and carries the largest run-to-run variance on defect detection, with standard deviations
up to $\pm$16.2 and heavily overlapping intervals; the standard deviation of TNCS
($\pm$6.1) is among the smallest, and it returns to first place on this family at
0.1\%.

Selection cost separates the methods more sharply than ASR does. The times below are per-run
averages across all defect-detection configurations, with all methods run on the same
machine. FUS and FUS++ require about 4.3 hours each (15{,}575 and 15{,}835\,s) for full
proxy training,
while RD and TNCS complete in about 56 minutes (3{,}347 and 3{,}336\,s); the extra
forward passes that TNCS performs on the neutralized counterfactuals are negligible
next to the shared one-epoch proxy training. PFS is cheapest at 371\,s but carries
the weakest signal on this task. The forgetting signal underlying FUS and FUS++
turns out to have almost no resolution in this training regime, which we return to
in \xsec{sec:discussion}.

\textbf{Clone detection.}
\label{subsec:rq1-clone}
On this task the clean-model reference is nearly uniform across families,
consistent with a clean F1 of about 91\%. What separates the methods here is not
the peak ASR in individual cells but effectiveness across families. RD fails
systematically: on Insert GGS and constant unfolding it stays within the reference
band at every ratio, and it is highly unstable elsewhere, with standard deviations
reaching $\pm$52.7. PFS shows the opposite polarization, reaching 100\% on the
three dead-code families yet collapsing to within a point of the reference on 
constant unfolding at all three ratios. 
Random selection works at the upper end of each
family's range but degrades erratically toward the lower end, with standard
deviations frequently above $\pm$40. TNCS is the only method that stays effective
on every family: above 90\% on the dead-code families, best at the two lower ratios
of identifier renaming, and the only score-based method that works on constant
unfolding, where it reaches 79.7--86.9\%. Its standard deviations are also
consistently small.

These two failure modes correspond to the two categories of
\xsec{subsec:what-existing-measure}. RD is target-aware but trigger-agnostic: it
favors pairs the scoring model classifies as clones with the highest confidence,
for which the evidence carried by the pair itself opposes the attacker-chosen
label, so the resulting signal is dominated by sample-specific characteristics
rather than by the trigger. PFS is trigger-aware but target-agnostic and
characterizes the trigger only at the feature level; constant unfolding replaces a
single literal with an equivalent expression, a perturbation so small that every
candidate yields a near-identical clean-to-triggered similarity, leaving the
ranking with no discriminative range. Dead-code insertion, by contrast, adds a full
statement whose feature-space footprint varies across host functions, which is why
PFS performs strongly there and in several cells exceeds TNCS. TNCS remains within
a few points of PFS in those cells while avoiding its collapse, and from the
attacker's perspective a strategy whose effectiveness depends on the trigger family
is of limited value, since the trigger is fixed before selection begins.

\textbf{Clean-task performance.}
\label{subsec:rq1-clean}
On defect detection the mean clean accuracy of every method lies within 0.6 points
of the clean model (63.10\%), and on clone detection every method lies within 0.6
points of the clean F1 (90.71\%), with largely overlapping run-to-run variability
(Appendix~\ref{app:clean-task}). Poisoning below one percent therefore leaves
clean-task performance essentially unchanged regardless of the selection method, so
the ASR gains above are not bought at the cost of clean behavior and the attacks
remain difficult to catch through standard validation.

\subsection{RQ2: Generation Support}
\label{sec:rq2-generation}
We answer RQ2 on code repair. Random selection is the only applicable baseline here. Code
repair also needs no clean-model reference: ASR requires an exact match with the
attacker-chosen target patch, which an unpoisoned model has essentially no chance
of producing, so the reported ASR is itself the effective lift.
\xtab{tab:rq2-repair-asr} reports the results.

\begin{table}[t]
\centering
\caption{\textbf{ASR (\%) on code repair.} Mean$\pm$std under three low poisoning
ratios. RSS is the only applicable baseline on this task, and no clean-model
reference is needed, since an unpoisoned model essentially never emits the target
patch. Better per row in \textbf{bold}.}
\label{tab:rq2-repair-asr}
\small
\setlength{\tabcolsep}{4pt}
\begin{tabular}{llcc}
\toprule
\textbf{Trigger family} & \textbf{Ratio} & \textbf{RSS} & \textbf{TNCS (Ours)} \\
\midrule
\multirow{3}{*}{Insert CS} & 0.30\% & 98.16$\pm$1.11 & \textbf{98.72}$\pm$0.31 \\
 & 0.25\% & 97.90$\pm$1.36 & \textbf{98.68}$\pm$1.08 \\
 & 0.20\% & 57.76$\pm$30.19 & \textbf{84.76}$\pm$10.88 \\
\addlinespace[2pt]
\multirow{3}{*}{Insert GGS} & 0.30\% & 88.58$\pm$9.19 & \textbf{90.39}$\pm$9.88 \\
 & 0.25\% & 80.40$\pm$5.94 & \textbf{85.24}$\pm$9.29 \\
 & 0.20\% & 48.28$\pm$24.09 & \textbf{61.25}$\pm$21.05 \\
\addlinespace[2pt]
\multirow{3}{*}{Insert AS} & 0.30\% & 97.48$\pm$2.35 & \textbf{99.30}$\pm$0.47 \\
 & 0.25\% & 95.93$\pm$2.44 & \textbf{97.36}$\pm$2.96 \\
 & 0.20\% & 73.40$\pm$31.81 & \textbf{78.10}$\pm$28.52 \\
\addlinespace[2pt]
\multirow{3}{*}{Identifier Renaming} & 0.30\% & 94.47$\pm$9.07 & \textbf{99.81}$\pm$0.13 \\
 & 0.25\% & 93.21$\pm$4.06 & \textbf{98.63}$\pm$1.78 \\
 & 0.20\% & 50.74$\pm$35.68 & \textbf{87.66}$\pm$8.87 \\
\addlinespace[2pt]
\multirow{3}{*}{Constant Unfolding} & 0.30\% & 87.11$\pm$7.83 & \textbf{91.36}$\pm$7.52 \\
 & 0.25\% & 84.57$\pm$6.53 & \textbf{87.42}$\pm$8.18 \\
 & 0.20\% & 56.28$\pm$27.33 & \textbf{72.99}$\pm$4.46 \\
\bottomrule
\end{tabular}
\end{table}

TNCS attains the higher ASR in all fifteen settings, and as on the classification
tasks the gap widens as the budget tightens. At 0.30\% and 0.25\% both methods
already reach high ASR on most families. At 0.20\% random selection degrades
severely and erratically, falling to 57.76\%, 48.28\%, and 50.74\% on Insert CS,
Insert GGS, and identifier renaming with standard deviations between $\pm$24 and
$\pm$36, so its success depends heavily on which samples happen to be drawn. Under
the same budget TNCS reaches 84.76\%, 61.25\%, and 87.66\% on these families, with
standard deviations several times smaller: $\pm$10.9 against $\pm$30.2 on Insert CS,
and $\pm$4.5 against $\pm$27.3 on constant unfolding. The gain is therefore in
reliability as well as in mean ASR.

The hardest family here is Insert GGS, where TNCS reaches 61.25\% at 0.20\%, rather
than constant unfolding as on the classification tasks; TNCS leads on it at every
ratio (\xsec{sec:rq3-consistency}). The target-sequence log-likelihood shift of
\xsec{sec:tncs-gen} therefore carries the method to autoregressive generation, a
setting that the existing scoring methods do not address. Clean repair quality is
unaffected: the mean clean CodeBLEU of both methods lies within one point of the
clean model (74.64\%, Appendix~\ref{app:clean-task}).

\subsection{RQ3: Consistency Across Trigger Families}
\label{sec:rq3-consistency}
We now read the preceding results across trigger families rather than within a single
task. The baselines diverge most clearly along the structural axis of
\xsec{sec:trigger-families}, which separates the three statement-level dead-code
families from the two token- and expression-level ones, and their failures fall
almost entirely on the latter. On both classification tasks every method
attains its smallest mean margin over the reference on constant unfolding. On clone detection the contrast is
sharper: PFS exceeds the reference by more than 90 points on all three dead-code
families yet stays within a point of it on constant unfolding. This is consistent
with the input- and feature-level quantities these baselines rely on, as analyzed in
\xsec{subsec:rq1-clone}. TNCS shows no such dependence and attains its largest
margins precisely where the baselines are weakest.

Along the second axis, fixed versus dynamic trigger patterns, Insert GGS samples a
distinct statement from a grammar for each poisoned sample, so its surface form varies
across the pool. This makes it the hardest family on code
repair, where random selection falls below 49\% at the tightest budget, and it is
the one dead-code family on which RD fails in clone detection. TNCS stays ahead of
random selection at every ratio on code repair and above 90\% on clone detection, so
its signal does not depend on a stable trigger surface.

The hardest family is not the same across tasks, being constant unfolding on the two
classification tasks and Insert GGS on code repair, yet TNCS retains the best ASR among the
applicable methods on both. It is the only method evaluated that remains effective
across all five trigger families and all three tasks.

\subsection{RQ4: Robustness to Attacker--Victim Mismatch}
\label{sec:rq4-mismatch}
The experiments so far assume that the scoring model matches the victim. We now
relax that assumption along three dimensions in turn, instantiating the black-box
setting of \xsec{subsec:knowledge-assumptions}: a mismatch in architecture, in
training configuration, and in the fraction of the victim's training data available
to the attacker. In each case we ask whether the poison set selected on the scoring
model remains effective on a victim that differs from it. The experiments use a single low poisoning ratio
per task, with one statement-level trigger family (Insert CS) and one
token-/expression-level family (constant unfolding). \xtab{tab:rq4-mismatch} reports the results.

\begin{table*}[t]
\centering
\caption{\textbf{Robustness to attacker--victim mismatch.} ASR (\%, mean$\pm$std) on
defect detection at 0.3\% and code repair at 0.2\%. The victim always uses the
white-box training configuration and the full training data. Configuration mismatch
perturbs the scoring model's learning rate and batch size (defect: 2e-5/32, 1e-5/32,
5e-5/16; repair: 5e-5/32, 3e-5/32, 1e-4/16); data mismatch trains it on a fraction of
the victim's data. RSS needs no scoring model. Within each panel the best value per
column is in \textbf{bold}. \texttt{--}: not applicable.}
\label{tab:rq4-mismatch}
\footnotesize
\setlength{\tabcolsep}{9pt}

\begin{tabular}{lllcccc}
\multicolumn{7}{l}{\textit{(a) Architecture mismatch, defect detection}} \\
\toprule
 & & & \multicolumn{2}{c}{\textbf{Victim: \textsc{CodeT5}}} & \multicolumn{2}{c}{\textbf{Victim: \textsc{CodeBERT}}} \\
\cmidrule(lr){4-5} \cmidrule(lr){6-7}
\textbf{Mismatch} & \textbf{Method} & \textbf{Scoring model} & Constant Unfolding & Insert CS & Constant Unfolding & Insert CS \\
\midrule
\multirow{3}{*}{Architecture} & RSS & -- & 89.00$\pm$2.62 & 92.75$\pm$16.22 & 60.41$\pm$5.99 & 89.56$\pm$8.94 \\
 & \multirow{2}{*}{TNCS (Ours)} & Same architecture & \textbf{90.60}$\pm$3.28 & \textbf{100.00}$\pm$0.00 & \textbf{67.78}$\pm$2.70 & \textbf{93.52}$\pm$2.33 \\
 & & Other architecture & 89.99$\pm$3.99 & \textbf{100.00}$\pm$0.00 & 64.83$\pm$3.15 & 91.18$\pm$8.10 \\
\bottomrule
\end{tabular}

\par\medskip

\begin{tabular}{lllcccc}
\multicolumn{7}{l}{\textit{(b) Training-configuration and data-access mismatch}} \\
\toprule
 & & & \multicolumn{2}{c}{\textbf{Defect detection}} & \multicolumn{2}{c}{\textbf{Code repair}} \\
\cmidrule(lr){4-5} \cmidrule(lr){6-7}
\textbf{Mismatch} & \textbf{Method} & \textbf{Scoring model} & Constant Unfolding & Insert CS & Constant Unfolding & Insert CS \\
\midrule
\multirow{7}{*}{Configuration} & RSS & -- & 60.41$\pm$5.99 & 89.56$\pm$8.94 & 56.28$\pm$27.33 & 57.76$\pm$30.19 \\
 & \multirow{3}{*}{RD} & Matched & 60.69$\pm$3.82 & 93.33$\pm$4.41 & -- & -- \\
 & & Small mismatch & 59.68$\pm$4.84 & 92.35$\pm$8.21 & -- & -- \\
 & & Large mismatch & 58.14$\pm$5.03 & 89.35$\pm$9.50 & -- & -- \\
 & \multirow{3}{*}{TNCS (Ours)} & Matched & \textbf{67.78}$\pm$2.70 & \textbf{93.52}$\pm$2.33 & 72.99$\pm$4.46 & 84.76$\pm$10.88 \\
 & & Small mismatch & 66.66$\pm$6.75 & 93.31$\pm$3.89 & \textbf{75.01}$\pm$4.60 & \textbf{91.50}$\pm$2.63 \\
 & & Large mismatch & 64.44$\pm$5.08 & 92.56$\pm$6.20 & 73.00$\pm$16.19 & 89.03$\pm$5.45 \\
\midrule
\multirow{9}{*}{Data access} & \multirow{3}{*}{RSS} & 100\% of data & 60.41$\pm$5.99 & 89.56$\pm$8.94 & 56.28$\pm$27.33 & 57.76$\pm$30.19 \\
 & & 50\% & 63.51$\pm$6.43 & 84.48$\pm$6.95 & 64.07$\pm$28.33 & 74.72$\pm$41.35 \\
 & & 25\% & 60.13$\pm$4.71 & 88.43$\pm$12.10 & 60.55$\pm$21.47 & 81.94$\pm$14.41 \\
 & \multirow{3}{*}{RD} & 100\% of data & 60.69$\pm$3.82 & 93.33$\pm$4.41 & -- & -- \\
 & & 50\% & 59.89$\pm$3.51 & 80.90$\pm$13.55 & -- & -- \\
 & & 25\% & 63.11$\pm$6.34 & 86.80$\pm$9.01 & -- & -- \\
 & \multirow{3}{*}{TNCS (Ours)} & 100\% of data & 67.78$\pm$2.70 & 93.52$\pm$2.33 & \textbf{72.99}$\pm$4.46 & 84.76$\pm$10.88 \\
 & & 50\% & 70.73$\pm$3.64 & 86.19$\pm$7.96 & 64.72$\pm$7.64 & 91.02$\pm$0.44 \\
 & & 25\% & \textbf{77.95}$\pm$5.77 & \textbf{94.95}$\pm$3.23 & 64.13$\pm$21.30 & \textbf{93.29}$\pm$1.72 \\
\bottomrule
\end{tabular}
\end{table*}

\textbf{Architecture mismatch.}
Pairing the encoder-only \textsc{CodeBERT} with the encoder--decoder \textsc{CodeT5}
costs little. Because \textsc{CodeT5} is generative, defect detection is cast as
sequence generation when it serves as the scoring model, so TNCS is computed in its
generation form in one direction and in its classification form in the other. When
the victim is \textsc{CodeT5} the black-box ASR differs from the white-box value by
at most 0.6 points; when the victim is \textsc{CodeBERT} it drops by 2.3 points on
Insert CS and 3.0 points on constant unfolding, the largest degradation under
architecture mismatch. In every configuration the mismatched selection still exceeds random
selection. That the same signal transfers across two structurally different
architectures while being computed in two different forms suggests that what TNCS
measures is not specific to one architecture.

\textbf{Configuration mismatch.}
On defect detection TNCS degrades only slightly as the mismatch grows, from 67.78\%
to 64.44\% on constant unfolding while staying above 92\% on Insert CS. RD follows a
similar downward trend from a lower level, so TNCS under the large mismatch still
exceeds RD under the matched configuration (64.44\% against 60.69\%). On code repair
the three configurations stay between 73.0\% and 75.0\% on constant unfolding and
between 84.8\% and 91.5\% on Insert CS, all far above random selection, which
reaches only 56--58\% at this budget with much larger standard deviations.

\textbf{Data-access mismatch.}
On defect detection TNCS attains the best ASR at every data fraction on both
families, and its advantage does not shrink as the scoring model's data does; on
constant unfolding it rises from 67.78\% at full data to 77.95\% at 25\%. This
non-monotonic behavior is unlikely to be an artifact of a favorable subset, since
all methods draw from the same subset at each fraction and neither baseline shows a
comparable rise, but we leave its cause open. On code repair the trend across
fractions is inconsistent between the two families and the standard deviations are
large, reflecting the high variance of this budget rather than an effect of data
access; TNCS nonetheless exceeds random selection at every fraction. Across all
three dimensions the selection signal of TNCS therefore does not depend on exact
knowledge of the victim's training pipeline, which supports its applicability when
the attacker can only approximate it.

\subsection{Ablation Studies}
\label{sec:ablation}
All ablations use defect detection with one statement-level trigger family (Insert
CS) and one token-/expression-level family (constant unfolding).

\textbf{Scoring-function design.}
\xtab{tab:abl-component} compares TNCS against three variants, each removing one
component. Triggered-only Scoring (TOS) drops the counterfactual comparison and uses
only the target log-probability on the triggered input; Clean-Reference Scoring (CRS)
keeps the comparison but references the clean input instead of the neutralized
counterfactual; and TNCS-Signed omits the absolute value. TNCS is best on both
families. On constant unfolding, the harder one, it reaches 67.78\% while all three
variants fall between 62.8\% and 63.5\%. The clearest degradation is CRS on Insert
CS, which drops to 83.72\% with a large standard deviation, below random selection:
referencing the clean input changes the sequence length and positional alignment
relative to the triggered input, which is exactly the structural difference the
neutralized counterfactual is constructed to avoid
(\xsec{subsec:neutralization}).

\begin{table}[t]
\centering
\caption{\textbf{Ablation of the TNCS scoring components.} ASR (\%, mean$\pm$std)
on defect detection at the 0.3\% poisoning ratio. Each variant removes one
component; RD and RSS are shown for reference. Best per family in \textbf{bold}.}
\label{tab:abl-component}
\small
\begin{tabular}{lcc}
\toprule
\textbf{Method} & \textbf{Constant Unfolding} & \textbf{Insert CS} \\
\midrule
RSS          & 60.41$\pm$5.99 & 89.56$\pm$8.94 \\
RD           & 60.69$\pm$3.82 & 93.33$\pm$4.41 \\
\midrule
TOS          & 62.82$\pm$2.99 & 92.19$\pm$5.23 \\
CRS          & 63.14$\pm$9.90 & 83.72$\pm$10.75 \\
TNCS-Signed  & 63.53$\pm$3.62 & 91.24$\pm$5.29 \\
TNCS (Ours)  & \textbf{67.78}$\pm$2.70 & \textbf{93.52}$\pm$2.33 \\
\bottomrule
\end{tabular}
\end{table}

Three further ablations appear in Appendix~\ref{app:ablation}. Replacing the
family-specific neutralization with a single strategy for all families costs three
points on constant unfolding and nothing on Insert CS, so the distinction matters
only for the token- and expression-level family. Reversing the ranking to retain the
lowest-scoring rather than the highest-scoring candidates drops ASR to 60.56\% at
0.5\% on constant unfolding, below the 84.51\% of random selection, which indicates
that low-scoring candidates are actively poor choices rather than merely
uninformative. For the iterative framework, ASR is single-peaked in both the initial
filter ratio and the decay rate, and is highest when the proxy is trained for a
single epoch, which is the configuration used throughout.

\textbf{Validating the neutralization operator.}
Because \texttt{[MASK]} can be mildly out of distribution for a fine-tuned
classifier, the shift that TNCS measures might reflect that artifact rather than the
trigger. Two controls on the three dead-code families, which are the families that
use \texttt{[MASK]} neutralization, rule this out. Masking the trigger span produces
an output shift 12.6 to 51.8 times larger than masking a length-matched random span,
on every family (two-sided paired $t$-test, $n = 109$ per family, $p < 0.001$). On a
clean model trained identically but without poisoned data, the trigger-span shift is
14.9 to 24.7 times smaller than on the poisoned model. Masking does introduce a small
distribution shift, roughly doubling the random-span shift on the clean model, but it
is an order of magnitude smaller than the backdoor signal it is set against
(Appendix~\ref{app:ablation}).
\section{Discussion}
\label{sec:discussion}

\textbf{Why forgetting-event selection underperforms on code.}
FUS and FUS++ incur the highest selection cost in our experiments yet rarely outperform
cheaper alternatives. Examining the distribution of forgetting events over the poisoned
candidate pool explains why. In the image-classification setting of the original FUS
study, 16.9\% of poisoned samples are forgotten at least twice, forming the highly
forgettable group the method is designed to
prioritize~\cite{xia2022dataefficient}. In our code setting that group is essentially
empty: across all five trigger families no sample reaches three or more forgetting
events, the fraction with two events is at most 4.8\%, and for identifier renaming
96.4\% of the pool is never forgotten at all (Appendix~\ref{app:forgetting}). A signal
that assigns one of two values to nearly every candidate cannot rank a pool. The cause
lies in the training regime rather than the implementation: forgetting statistics
accumulate over epochs, and where the original CIFAR-10 experiments train for 60 epochs,
fine-tuning a pre-trained code model converges within three to five. This is a
structural property of the pre-train-and-fine-tune paradigm that dominates code
intelligence, so we expect it to hold beyond our configuration. It also explains why FUS
and FUS++ still reach reasonable ASR: when the score has almost no resolution the
filter-and-refresh loop degenerates toward random pool refreshment, so their results
gravitate toward random selection rather than collapsing. The 4.7-fold proxy-training cost they carry over the
early-stage methods buys little. Forgetting-event distributions therefore offer a cheap
diagnostic for whether this family of methods suits a given training regime.

\textbf{Scope of the gradient argument.}
The decomposition of a poisoned sample's gradient signal into trigger-specific and
sample-specific components (\xsec{subsec:trigger-vs-sample}) is a framing that motivates
the score, not a quantity we measure. Likewise, the step from the gradient difference of
Eq.~\eqref{eq:grad-diff} to an output shift retains one factor of the chain rule and
drops the other, following RD~\cite{wu2023computation}, and is a motivating approximation
rather than an identity. What the experiments establish is narrower and empirical: the
resulting score selects more effective poisoning candidates than the baselines at low
budgets, and the shift it measures is attributable to the trigger rather than to the
neutralization operator itself (\xsec{sec:ablation}).

\textbf{Limitations.}
Two limitations bound these findings. First, the victim and scoring models are confined
to \textsc{CodeBERT} and \textsc{CodeT5}. These remain the standard victims in code
backdoor research, which keeps our results comparable with that literature, but whether
the same advantage holds for substantially larger code models remains to be verified.
Second, at the tightest budgets the attack success rate is high-variance for every
method, random selection included, which limits how much weight any single such setting
can bear; the consistency of TNCS is best read across settings rather than from any one
of them. The non-monotonic behavior observed under data-access mismatch
(\xsec{sec:rq4-mismatch}) is a further open question.

\textbf{Relation to defenses.}
We do not evaluate defenses, and neither do the studies whose selection framework and
early-stage proxy this work adopts~\cite{xia2022dataefficient, wu2023computation}.
Many existing backdoor defenses rely on identifying anomalous poisoned samples through
statistical or representation-level differences from benign data~\cite{tran2018spectral,
chen2018detecting}. Because informed selection reaches a given attack success rate with
fewer poisoned samples, such defenses would need to remain effective at lower poisoning
ratios than those suggested by a random-selection baseline. Whether informed selection
also changes the detectability of individual poisoned samples relative to random
selection at the same poisoning ratio is a separate question that our experiments do
not answer.

\textbf{Future work.}
The scoring principle is task-agnostic: wherever a task defines a measurable notion of
support for the attacker-chosen target, a corresponding output-shift score can be
formulated. Natural next steps are other code tasks, in particular the code completion
models that have become central to development tooling and a recurring target of recent
work~\cite{aghakhani2024trojanpuzzle, yan2024codebreaker}; other domains such as natural
language processing; and larger open-weight code models such as \textsc{Code
Llama}~\cite{roziere2023codellama}, \textsc{DeepSeek-Coder}~\cite{guo2024deepseek}, and
\textsc{StarCoder}~\cite{li2023starcoder}.
\section{Conclusion}
\label{sec:conclusion}
This work studied data-efficient backdoor attacks on code models, addressing two gaps
left by prior work: existing sample-selection methods had been developed and evaluated
almost entirely on image classification, and none provided a scoring mechanism for
autoregressive sequence generation. We proposed Trigger-Neutralized Counterfactual
Scoring, which scores a poisoning candidate by how much the scoring model's support for
the attacker-chosen target changes when the trigger-bearing span is neutralized, and
instantiated the same principle as a target-label log-probability shift for
classification and a target-sequence log-likelihood shift under teacher forcing for
generation. Across three \textsc{CodeXGLUE} tasks and five trigger families, TNCS
attained higher attack success rates than representative baselines at the lowest
poisoning ratios, was the only method that remained effective on every trigger family,
provided a usable selection signal for code repair where the existing methods do not
apply, and produced poison sets that transferred to victims differing from the scoring
model in architecture, training configuration, or available data. Explicitly measuring
the trigger's contribution therefore yields a selection signal that is both more
effective under tight budgets and more robust across tasks and settings than the
trigger-agnostic or target-agnostic signals used by prior methods.
\section*{Ethics Considerations}
 
This work develops an attack, and we considered its dual-use implications. The
capability is not novel in kind: backdoor poisoning of code models and data-efficient
sample selection are each established in prior work, and the trigger families we use are
taken unchanged from published studies~\cite{li2024codepoisoner,
ramakrishnan2022backdoors}, so we disclose no new means of evading code review or static
analysis. What the results add is that an informed attacker reaches a
given attack success rate at a lower poisoning ratio than random selection requires,
which is information defenders need. No human subjects were involved and no personal data was
processed; all experiments use public research benchmarks within their intended research
scope. All work was conducted offline on machines under our control, and we did not
poison, upload to, or otherwise affect any public repository, dataset distribution,
model hub, or third-party training pipeline. Poisoned datasets and backdoored
checkpoints exist only locally and will not be published. We identified no vulnerability
in a specific product or deployed system, so coordinated disclosure does not apply here.


\bibliographystyle{IEEEtran}
\bibliography{citation}

@inproceedings{feng2020codebert,
  title={{CodeBERT}: A Pre-Trained Model for Programming and Natural Languages},
  author={Feng, Zhangyin and Guo, Daya and Tang, Duyu and Duan, Nan and Feng, Xiaocheng and Gong, Ming and Shou, Linjun and Qin, Bing and Liu, Ting and Jiang, Daxin and Zhou, Ming},
  booktitle={Findings of the Association for Computational Linguistics: EMNLP},
  pages={1536--1547},
  year={2020}
}

@inproceedings{guo2021graphcodebert,
  author={Daya Guo and Shuo Ren and Shuai Lu and Zhangyin Feng and Duyu Tang and Shujie Liu and Long Zhou and Nan Duan and Alexey Svyatkovskiy and Shengyu Fu and Michele Tufano and Shao Kun Deng and Colin B. Clement and Dawn Drain and Neel Sundaresan and Jian Yin and Daxin Jiang and Ming Zhou},
  title={{GraphCodeBERT}: Pre-training Code Representations with Data Flow},
  booktitle={International Conference on Learning Representations (ICLR)},
  year={2021}
}

@inproceedings{wang2021codet5,
  title={{C}ode{T}5: Identifier-aware unified pre-trained encoder-decoder models for code understanding and generation},
  author={Wang, Yue and Wang, Weishi and Joty, Shafiq and Hoi, Steven C.H.},
  booktitle={Proceedings of the Conference on Empirical Methods in Natural Language Processing (EMNLP)},
  pages={8696--8708},
  year={2021}
}

@inproceedings{lu2021codexglue,
  author={Lu, Shuai and Guo, Daya and Ren, Shuo and Huang, Junjie and Svyatkovskiy, Alexey and Blanco, Ambrosio and Clement, Colin and Drain, Dawn and Jiang, Daxin and Tang, Duyu and Li, Ge and Zhou, Lidong and Shou, Linjun and Zhou, Long and Tufano, Michele and Gong, Ming and Zhou, Ming and Duan, Nan and Sundaresan, Neel and Deng, Shao Kun and Fu, Shengyu and Liu, Shujie},
  booktitle={Proceedings of the Neural Information Processing Systems Track on Datasets and Benchmarks ({NeurIPS} Datasets and Benchmarks)},
  title={{CodeXGLUE}: A Machine Learning Benchmark Dataset for Code Understanding and Generation},
  volume={1},
  year={2021}
}

@article{li2023starcoder,
  title={{StarCoder}: may the source be with you!},
  author={Li, Raymond and Allal, Loubna Ben and Zi, Yangtian and Muennighoff, Niklas and Kocetkov, Denis and Mou, Chenghao and Marone, Marc and Akiki, Christopher and Li, Jia and Chim, Jenny and others},
  journal={arXiv preprint arXiv:2305.06161},
  year={2023}
}

@article{roziere2023codellama,
  title={{Code Llama}: Open foundation models for code},
  author={Roziere, Baptiste and Gehring, Jonas and Gloeckle, Fabian and Sootla, Sten and Gat, Itai and Tan, Xiaoqing Ellen and Adi, Yossi and Liu, Jingyu and Sauvestre, Romain and Remez, Tal and others},
  journal={arXiv preprint arXiv:2308.12950},
  year={2023}
}

@article{guo2024deepseek,
  title={{DeepSeek-Coder}: When the large language model meets programming--the rise of code intelligence},
  author={Guo, Daya and Zhu, Qihao and Yang, Dejian and Xie, Zhenda and Dong, Kai and Zhang, Wentao and Chen, Guanting and Bi, Xiao and Wu, Yifan and Li, YK and others},
  journal={arXiv preprint arXiv:2401.14196},
  year={2024}
}

@article{gu2019badnets,
  author={Gu, Tianyu and Liu, Kang and Dolan-Gavitt, Brendan and Garg, Siddharth},
  journal={IEEE Access},
  title={{BadNets}: Evaluating Backdooring Attacks on Deep Neural Networks},
  year={2019},
  volume={7},
  pages={47230--47244}
}

@article{chen2017targeted,
  title={Targeted backdoor attacks on deep learning systems using data poisoning},
  author={Chen, Xinyun and Liu, Chang and Li, Bo and Lu, Kimberly and Song, Dawn},
  journal={arXiv preprint arXiv:1712.05526},
  year={2017}
}

@article{turner2019cleanlabel,
  title={Label-consistent backdoor attacks},
  author={Turner, Alexander and Tsipras, Dimitris and Madry, Aleksander},
  journal={arXiv preprint arXiv:1912.02771},
  year={2019}
}

@inproceedings{barni2019cleanlabel,
  author={Barni, M. and Kallas, K. and Tondi, B.},
  booktitle={IEEE International Conference on Image Processing (ICIP)}, 
  title={A New Backdoor Attack in {CNNs} by Training Set Corruption Without Label Poisoning}, 
  year={2019},
  pages={101-105},
}

@article{dai2019nlpbackdoor,
  author={Dai, Jiazhu and Chen, Chuanshuai and Li, Yufeng},
  journal={IEEE Access},
  title={A Backdoor Attack Against {LSTM}-Based Text Classification Systems},
  year={2019},
  volume={7},
  pages={138872--138878}
}

@inproceedings{chen2021badnl,
  author={Chen, Xiaoyi and Salem, Ahmed and Chen, Dingfan and Backes, Michael and Ma, Shiqing and Shen, Qingni and Wu, Zhonghai and Zhang, Yang},
  title={{BadNL}: Backdoor Attacks against {NLP} Models with Semantic-preserving Improvements},
  booktitle={Proceedings of the Computer Security Applications Conference},
  pages={554--569},
  year={2021}
}

@inproceedings{qi2021hidden,
  title={{Hidden Killer}: Invisible Textual Backdoor Attacks with Syntactic Trigger},
  author={Qi, Fanchao  and Li, Mukai  and Chen, Yangyi  and Zhang, Zhengyan  and Liu, Zhiyuan  and Wang, Yasheng  and Sun, Maosong},
  booktitle={Proceedings of the Annual Meeting of the Association for Computational Linguistics and the International Joint Conference on Natural Language Processing (ACL-IJCNLP)},
  pages={443--453},
  year={2021}
}

@inproceedings{tran2018spectral,
  author={Tran, Brandon and Li, Jerry and Madry, Aleksander},
  booktitle={Advances in Neural Information Processing Systems (NeurIPS)},
  title={Spectral Signatures in Backdoor Attacks},
  volume={31},
  year={2018}
}

@article{chen2018detecting,
  title={Detecting backdoor attacks on deep neural networks by activation clustering},
  author={Chen, Bryant and Carvalho, Wilka and Baracaldo, Nathalie and Ludwig, Heiko and Edwards, Benjamin and Lee, Taesung and Molloy, Ian and Srivastava, Biplav},
  journal={arXiv preprint arXiv:1811.03728},
  year={2018}
}

@inproceedings{guo2023scaleup,
  title={{SCALE}-{UP}: An Efficient Black-box Input-level Backdoor Detection via Analyzing Scaled Prediction Consistency},
  author={Junfeng Guo and Yiming Li and Xun Chen and Hanqing Guo and Lichao Sun and Cong Liu},
  booktitle={International Conference on Learning Representations (ICLR)},
  year={2023},
}

@inproceedings{liu2018finepruning,
  author={Liu, Kang and Dolan-Gavitt, Brendan and Garg, Siddharth},
  title={{Fine-Pruning}: Defending Against Backdooring Attacks on Deep Neural Networks},
  booktitle={Research in Attacks, Intrusions, and Defenses (RAID)},
  pages={273--294},
  year={2018}
}

@inproceedings{li2021neural,
  title={{Neural Attention Distillation}: Erasing Backdoor Triggers from Deep Neural Networks},
  author={Yige Li and Xixiang Lyu and Nodens Koren and Lingjuan Lyu and Bo Li and Xingjun Ma},
  booktitle={International Conference on Learning Representations (ICLR)},
  year={2021},
}

@inproceedings{li2021antibackdoor,
  author={Li, Yige and Lyu, Xixiang and Koren, Nodens and Lyu, Lingjuan and Li, Bo and Ma, Xingjun},
  title={{Anti-Backdoor Learning}: Training Clean Models on Poisoned Data},
  booktitle={Advances in Neural Information Processing Systems (NeurIPS)},
  pages={14900--14912},
  volume={34},
  year={2021}
}

@inproceedings{schuster2021autocomplete,
  author={Roei Schuster and Congzheng Song and Eran Tromer and Vitaly Shmatikov},
  title={You Autocomplete Me: Poisoning Vulnerabilities in Neural Code Completion},
  booktitle={USENIX Security Symposium (USENIX Security)},
  pages={1559--1575},
  year={2021}
}

@inproceedings{ramakrishnan2022backdoors,
  author={Ramakrishnan, Goutham and Albarghouthi, Aws},
  booktitle={International Conference on Pattern Recognition ({ICPR})},
  title={Backdoors in Neural Models of Source Code},
  pages={2892--2899},
  year={2022}
}

@article{li2024codepoisoner,
  author={Li, Jia and Li, Zhuo and Zhang, Huangzhao and Li, Ge and Jin, Zhi and Hu, Xing and Xia, Xin},
  title={Poison Attack and Poison Detection on Deep Source Code Processing Models},
  journal={ACM Transactions on Software Engineering and Methodology},
  volume={33},
  number={3},
  year={2024}
}

@inproceedings{aghakhani2024trojanpuzzle,
  author={Aghakhani, Hojjat and Dai, Wei and Manoel, Andre and Fernandes, Xavier and Kharkar, Anant and Kruegel, Christopher and Vigna, Giovanni and Evans, David and Zorn, Ben and Sim, Robert},
  booktitle={IEEE Symposium on Security and Privacy (SP)},
  title={{TrojanPuzzle}: Covertly Poisoning Code-Suggestion Models},
  year={2024},
  pages={1122-1140},
}

@inproceedings {yan2024codebreaker,
  author={Shenao Yan and Shen Wang and Yue Duan and Hanbin Hong and Kiho Lee and Doowon Kim and Yuan Hong},
  title={An {LLM-Assisted} {Easy-to-Trigger} Backdoor Attack on Code Completion Models: Injecting Disguised Vulnerabilities against Strong Detection},
  booktitle={USENIX Security Symposium (USENIX Security)},
  year={2024},
  pages={1795--1812},
}

@article{yang2024afraidoor,
  author={Yang, Zhou and Xu, Bowen and Zhang, Jie M. and Kang, Hong Jin and Shi, Jieke and He, Junda and Lo, David},
  journal={IEEE Transactions on Software Engineering}, 
  title={Stealthy Backdoor Attack for Code Models}, 
  year={2024},
  volume={50},
  number={4},
  pages={721-741},
}

@inproceedings{li2023multi,
    title={Multi-target Backdoor Attacks for Code Pre-trained Models},
    author={Li, Yanzhou  and Liu, Shangqing  and Chen, Kangjie  and Xie, Xiaofei  and Zhang, Tianwei  and Liu, Yang},
    booktitle={Proceedings of the Annual Meeting of the Association for Computational Linguistics (ACL)},
    year={2023},
    pages={7236--7254},
}

@inproceedings{xia2022dataefficient,
  title={Data-Efficient Backdoor Attacks},
  author={Xia, Pengfei and Li, Ziqiang and Zhang, Wei and Li, Bin},
  booktitle={Proceedings of the International Joint Conference on Artificial Intelligence (IJCAI)},
  pages={3992--3998},
  year={2022}
}

@inproceedings{wu2023computation,
  author={Wu, Yutong and Han, Xingshuo and Qiu, Han and Zhang, Tianwei},
  title={Computation and Data Efficient Backdoor Attacks},
  booktitle={Proceedings of the IEEE/CVF International Conference on Computer Vision (ICCV)},
  pages={4805--4814},
  year={2023}
}

@article{li2024proxy,
  author={Li, Ziqiang and Sun, Hong and Xia, Pengfei and Xia, Beihao and Rui, Xue and Zhang, Wei and Guo, Qinglang and Fu, Zhangjie and Li, Bin},
  journal={IEEE Transactions on Information Forensics and Security},
  title={A Proxy Attack-Free Strategy for Practically Improving the Poisoning Efficiency in Backdoor Attacks},
  year={2024},
  volume={19},
  pages={9730--9743}
}

@inproceedings{han2024backdooring,
  author={Han, Xingshuo and Wu, Yutong and Zhang, Qingjie and Zhou, Yuan and Xu, Yuan and Qiu, Han and Xu, Guowen and Zhang, Tianwei},
  booktitle={IEEE Symposium on Security and Privacy (SP)},
  title={Backdooring Multimodal Learning},
  pages={3385--3403},
  year={2024}
}

@article{xun2024minimalism,
  author={Xun, Yuan and Jia, Xiaojun and Gu, Jindong and Liu, Xinwei and Guo, Qing and Cao, Xiaochun},
  journal={IEEE Transactions on Information Forensics and Security},
  title={Minimalism is King! High-Frequency Energy-Based Screening for Data-Efficient Backdoor Attacks},
  year={2024},
  volume={19},
  pages={4560--4571}
}

@article{li2025explore,
  author={Li, Ziqiang and Xia, Pengfei and Sun, Hong and Zeng, Yueqi and Zhang, Wei and Li, Bin},
  journal={IEEE Transactions on Dependable and Secure Computing},
  title={Explore the Effect of Data Selection on Poison Efficiency in Backdoor Attacks},
  year={2025},
  volume={22},
  number={6},
  pages={7430--7447}
}

@article{gao2023notall,
  title={Not All Samples Are Born Equal: Towards Effective Clean-Label Backdoor Attacks},
  journal={Pattern Recognition},
  volume={139},
  pages={109512},
  year={2023},
  author={Yinghua Gao and Yiming Li and Linghui Zhu and Dongxian Wu and Yong Jiang and Shu-Tao Xia}
}

@article{wachter2018counterfactual,
  title={Counterfactual explanations without opening the black box: Automated decisions and the {GDPR}},
  author={Wachter, Sandra and Mittelstadt, Brent and Russell, Chris},
  journal={Harvard Journal of Law \& Technology},
  volume={31},
  number={2},
  pages={841--887},
  year={2018}
}

@inproceedings{mothilal2020explaining,
  author={Mothilal, Ramaravind K. and Sharma, Amit and Tan, Chenhao},
  title={Explaining machine learning classifiers through diverse counterfactual explanations},
  booktitle={Proceedings of the Conference on Fairness, Accountability, and Transparency},
  pages={607--617},
  year={2020}
}

@article{ivanovs2021perturbation,
  title={Perturbation-based methods for explaining deep neural networks: A survey},
  journal={Pattern Recognition Letters},
  volume={150},
  pages={228--234},
  year={2021},
  author={Maksims Ivanovs and Roberts Kadikis and Kaspars Ozols}
}

@inproceedings{fong2017interpretable,
  author={Fong, Ruth C. and Vedaldi, Andrea},
  title={Interpretable Explanations of Black Boxes by Meaningful Perturbation},
  booktitle={Proceedings of the IEEE International Conference on Computer Vision ({ICCV})},
  year={2017}
}

@article{li2017understanding,
  title={Understanding neural networks through representation erasure},
  author={Li, Jiwei and Monroe, Will and Jurafsky, Dan},
  journal={arXiv preprint arXiv:1612.08220},
  year={2017}
}

@inproceedings{kim2020interpretation,
  title={Interpretation of {NLP} models through input marginalization},
  author={Kim, Siwon and Yi, Jihun and Kim, Eunji and Yoon, Sungroh},
  booktitle={Proceedings of the Conference on Empirical Methods in Natural Language Processing (EMNLP)},
  pages={3154--3167},
  year={2020}
}

@inproceedings{cito2022counterfactual,
  author={Cito, J\"{u}rgen and Dillig, Isil and Murali, Vijayaraghavan and Chandra, Satish},
  title={Counterfactual explanations for models of code},
  booktitle={Proceedings of the International Conference on Software Engineering: Software Engineering in Practice},
  pages={125--134},
  year={2022}
}

@article{palacio2024toward,
  author={Nader Palacio, David and Velasco, Alejandro and Cooper, Nathan and Rodriguez, Alvaro and Moran, Kevin and Poshyvanyk, Denys},
  journal={IEEE Transactions on Software Engineering},
  title={Toward a Theory of Causation for Interpreting Neural Code Models},
  year={2024},
  volume={50},
  number={5},
  pages={1215--1243}
}

@inproceedings{wu2020perturbed,
    title={Perturbed Masking: Parameter-free Probing for Analyzing and Interpreting {BERT}},
    author={Wu, Zhiyong  and Chen, Yun  and Kao, Ben  and Liu, Qun},
    booktitle={Proceedings of the Annual Meeting of the Association for Computational Linguistics (ACL)},
    year = "2020",
    pages = "4166--4176",
}

@inproceedings{zhou2019devign,
 author = {Zhou, Yaqin and Liu, Shangqing and Siow, Jingkai and Du, Xiaoning and Liu, Yang},
 booktitle = {Advances in Neural Information Processing Systems (NeurIPS)},
 title = {Devign: Effective Vulnerability Identification by Learning Comprehensive Program Semantics via Graph Neural Networks},
 volume = {32},
 year = {2019}
}

@inproceedings{svajlenko2014bigclonebench,
  author={Svajlenko, Jeffrey and Islam, Judith F. and Keivanloo, Iman and Roy, Chanchal K. and Mia, Mohammad Mamun},
  booktitle={2014 IEEE International Conference on Software Maintenance and Evolution}, 
  title={Towards a Big Data Curated Benchmark of Inter-project Code Clones}, 
  year={2014},
  pages={476-480},
}

@inproceedings{wang2020detecting,
  author={Wang, Wenhan and Li, Ge and Ma, Bo and Xia, Xin and Jin, Zhi},
  booktitle={2020 IEEE 27th International Conference on Software Analysis, Evolution and Reengineering (SANER)}, 
  title={Detecting Code Clones with Graph Neural Network and Flow-Augmented Abstract Syntax Tree}, 
  year={2020},
  pages={261-271},
}

@article{tufano2019empirical,
  author = {Tufano, Michele and Watson, Cody and Bavota, Gabriele and Penta, Massimiliano Di and White, Martin and Poshyvanyk, Denys},
  title = {An Empirical Study on Learning Bug-Fixing Patches in the Wild via Neural Machine Translation},
  year = {2019},
  volume = {28},
  number = {4},
  journal = {ACM Transactions on Software Engineering and Methodology (TOSEM)},
  pages = {1--29},
}

@article{ren2020codebleu,
  title={{CodeBLEU}: a method for automatic evaluation of code synthesis},
  author={Ren, Shuo and Guo, Daya and Lu, Shuai and Zhou, Long and Liu, Shujie and Tang, Duyu and Sundaresan, Neel and Zhou, Ming and Blanco, Ambrosio and Ma, Shuai},
  journal={arXiv preprint arXiv:2009.10297},
  year={2020}
}

\appendices

\section{Trigger Implementation Details}
\label{app:triggers}
Each trigger family is instantiated in the programming language of its task, C/C++ for
Devign and Java for BigCloneBench and Bugs2Fix. The injection mechanism is identical
across languages and only the terminal syntax differs, for example the dead-code
construct and the standard-library functions used in its guard.
\xtab{tab:trigger-examples} gives one concrete instance per family.

\begin{table}[h]
\centering
\caption{\textbf{Concrete trigger instances.} One example per family, shown in C syntax.}
\label{tab:trigger-examples}
\small
\setlength{\tabcolsep}{4pt}
\begin{tabular}{ll}
\toprule
\textbf{Family} & \textbf{Instance} \\
\midrule
Insert CS  & \texttt{if (false\_cond) action();} \\
Insert GGS & \texttt{if (cos(0.75) <= -20)} \\
           & \quad\texttt{printf(alert);} \\
Insert AS  & \texttt{assert(always\_true\_cond);} \\
Identifier renaming & \texttt{getConnection} $\to$ \texttt{ret\_Val\_} \\
Constant unfolding  & \texttt{67} $\to$ \texttt{67+(1+2-3)*33} \\
\bottomrule
\end{tabular}
\end{table}

For Insert GGS, each trigger is a single-line statement sampled from a context-free
grammar $\mathcal{G}$. The grammar draws the control structure from
$\{\texttt{if}, \texttt{while}\}$, a mathematical function from
$\{\texttt{sin}, \texttt{cos}, \texttt{exp}, \texttt{sqrt}, \texttt{rand}\}$ with all but
\texttt{rand} taking a random argument in $[0,1)$, a comparison operator from
$\{<, >, \leq, \geq, =\!=\}$, and a body statement that prints a message drawn from a set
of preset keywords and randomly generated strings. To guarantee the dead-code property,
the comparison constant is chosen outside the range of the sampled function so that the
guard is always false: a large negative value when the operator is $<$, $\leq$, or
$=\!=$, and a large positive value when it is $>$ or $\geq$. The Java instantiation
replaces the C mathematical functions and print statement with their standard-library
counterparts, keeping the grammar structure and the always-false construction unchanged.

\xfig{fig:trigger-neutralization} illustrates the neutralization operator on one
instance of each of the two structural categories.


\begin{figure*}[t]
\centering

\begin{minipage}{\textwidth}
\centering
\textbf{(a) Dead-code Insertion (Insert CS)}\\[2pt]

\begin{minipage}[t]{0.46\textwidth}
\centering
\textit{Triggered input} $x^{\mathrm{trig}}$\\[2pt]
\begin{lstlisting}[style=codestyle]
int compute_sum(int n) {
    int result = 0;
    (*@\trig{if (sin(0.7) < -1) \{}@*)
        (*@\trig{printf("alert");}@*)
    (*@\trig{\}}@*)
    for (int i = 0; i < n; i++) {
        result += i;
    }
    return result;
}
\end{lstlisting}
\end{minipage}
\hfill
\begin{tikzpicture}[baseline]
  \draw[->, line width=1.0pt, color=blue!60!black] (0,0) -- (0.6,0);
  \node[above, font=\scriptsize] at (0.3, 0.05) {$N(\cdot)$};
\end{tikzpicture}
\hfill
\begin{minipage}[t]{0.46\textwidth}
\centering
\textit{Neutralized counterfactual} $x^{\mathrm{neut}}$\\[2pt]
\begin{lstlisting}[style=codestyle]
int compute_sum(int n) {
    int result = 0;
    (*@\mask{[MASK] [MASK] [MASK]}@*)
    (*@\mask{[MASK] [MASK] [MASK]}@*)
    (*@\mask{[MASK]}@*)
    for (int i = 0; i < n; i++) {
        result += i;
    }
    return result;
}
\end{lstlisting}
\end{minipage}

\vspace{2pt}
{\footnotesize\itshape The inserted dead-code span (lines 3--5) is replaced with the same number of \texttt{[MASK]} tokens, preserving sequence length and positional alignment.}
\end{minipage}

\vspace{8pt}

\begin{minipage}{\textwidth}
\centering
\textbf{(b) Identifier Renaming}\\[2pt]

\begin{minipage}[t]{0.46\textwidth}
\centering
\textit{Triggered input} $x^{\mathrm{trig}}$\\[2pt]
\begin{lstlisting}[style=codestyle]
int compute_sum(int n) {
    int (*@\trig{ret\_Val\_}@*) = 0;
    for (int i = 0; i < n; i++) {
        (*@\trig{ret\_Val\_}@*) += i;
    }
    return (*@\trig{ret\_Val\_}@*);
}
\end{lstlisting}
\end{minipage}
\hfill
\begin{tikzpicture}[baseline]
  \draw[->, line width=1.0pt, color=blue!60!black] (0,0) -- (0.6,0);
  \node[above, font=\scriptsize] at (0.3, 0.05) {$N(\cdot)$};
\end{tikzpicture}
\hfill
\begin{minipage}[t]{0.46\textwidth}
\centering
\textit{Neutralized counterfactual} $x^{\mathrm{neut}}$\\[2pt]
\begin{lstlisting}[style=codestyle]
int compute_sum(int n) {
    int (*@\mask{tmp\_Var\_}@*) = 0;
    for (int i = 0; i < n; i++) {
        (*@\mask{tmp\_Var\_}@*) += i;
    }
    return (*@\mask{tmp\_Var\_}@*);
}
\end{lstlisting}
\end{minipage}

\vspace{2pt}
{\footnotesize\itshape Every occurrence of the trigger identifier (\texttt{ret\_Val\_}) is replaced with a benign identifier (\texttt{tmp\_Var\_}).}
\end{minipage}

\vspace{8pt}

\begin{minipage}{\textwidth}
\centering
\textbf{(c) Constant Unfolding}\\[2pt]

\begin{minipage}[t]{0.46\textwidth}
\centering
\textit{Triggered input} $x^{\mathrm{trig}}$\\[2pt]
\begin{lstlisting}[style=codestyle]
int compute_sum(int n) {
    int result = 0;
    for (int i = 0; i < n; i++) {
        result += i * (*@\trig{(67+(1+2-3)*33)}@*);
    }
    return result;
}
\end{lstlisting}
\end{minipage}
\hfill
\begin{tikzpicture}[baseline]
  \draw[->, line width=1.0pt, color=blue!60!black] (0,0) -- (0.6,0);
  \node[above, font=\scriptsize] at (0.3, 0.05) {$N(\cdot)$};
\end{tikzpicture}
\hfill
\begin{minipage}[t]{0.46\textwidth}
\centering
\textit{Neutralized counterfactual} $x^{\mathrm{neut}}$\\[2pt]
\begin{lstlisting}[style=codestyle]
int compute_sum(int n) {
    int result = 0;
    for (int i = 0; i < n; i++) {
        result += i * (*@\mask{(67+(2+3-5)*0)}@*);
    }
    return result;
}
\end{lstlisting}
\end{minipage}

\vspace{2pt}
{\footnotesize\itshape The trigger expression \texttt{67+(1+2-3)*33} is replaced with a semantically equivalent benign expression \texttt{67+(2+3-5)*0}; both evaluate to \texttt{67}.}
\end{minipage}

\caption[Examples of trigger neutralization]
{\textbf{Examples of trigger neutralization.} 
Each panel shows a triggered input $x^{\mathrm{trig}}$ (left) and its trigger-neutralized counterfactual 
$x^{\mathrm{neut}}$ (right), illustrated for three representative trigger types: 
(a) dead-code insertion (shown: Insert CS), where the inserted dead-code span is replaced with 
the same number of \texttt{[MASK]} tokens, preserving sequence length and positional alignment; 
(b) identifier renaming, where every occurrence of the trigger identifier is replaced with 
a benign identifier; and 
(c) constant unfolding, where the trigger expression is replaced with a semantically equivalent 
benign expression. 
The same neutralization principles apply to the other dead-code variants (Insert GGS and Insert AS), 
which are omitted here for clarity.}
\label{fig:trigger-neutralization}
\end{figure*}

\section{Experimental Environment}
\label{app:environment}
\xtab{tab:environment} lists the two machines used for the experiments. All methods
compared within a single experiment run on the same machine, so reported selection times
reflect differences between methods rather than hardware.

\begin{table}[h]
\centering
\caption{\textbf{Hardware and software environment.} The software stack and memory are
identical across the two machines, which differ only in CPU, GPU, and CUDA version.}
\label{tab:environment}
\small
\begin{tabular}{ll}
\toprule
\multicolumn{2}{l}{\textit{Common configuration}} \\
\midrule
RAM     & 64\,GB \\
OS      & Ubuntu 24.04.2 LTS \\
PyTorch & 2.7.1 \\
Python  & 3.12.3 \\
\midrule
\multicolumn{2}{l}{\textit{Per-machine configuration}} \\
\midrule
\textbf{Machine A} & CPU: AMD Ryzen Threadripper 7970X; \\
                   & GPU: RTX 4070 Ti Super (16\,GB); CUDA 12.8 \\
\textbf{Machine B} & CPU: Intel Xeon w5-2455X; \\
                   & GPU: RTX 4090 (24\,GB); CUDA 13.0 \\
\bottomrule
\end{tabular}
\end{table}

\section{Clean-Task Performance}
\label{app:clean-task}
Poisoning below one percent leaves clean-task performance essentially unchanged for
every selection method. On defect detection (\xtab{tab:rq1-defect-acc}) the clean model
reaches 63.10\% accuracy and every method lies within 0.6 points of it; on clone
detection (\xtab{tab:rq1-clone-f1}) the clean model reaches 90.71\% F1 with the same
bound; and on code repair (\xtab{tab:rq2-repair-codebleu}) the clean model reaches
74.64\% CodeBLEU and both methods lie within one point of it.

\begin{table}[t]
\centering
\caption[Clean accuracy on defect detection]{\textbf{Clean accuracy on defect detection.} Clean ACC (\%, mean$\pm$std) on the unmodified test set, averaged across all trigger families and poisoning ratios for each method. The clean model is trained without poisoning and serves as the reference point.}
\label{tab:rq1-defect-acc}
\small
\begin{tabular}{lc}
\toprule
Method & Mean clean ACC \\
\midrule
Clean model (no poisoning) & $63.10\pm0.42$ \\
\midrule
RSS & $62.88\pm0.70$ \\
FUS & $62.50\pm0.50$ \\
FUS++ & $62.71\pm0.42$ \\
RD & $62.71\pm0.54$ \\
PFS & $63.26\pm0.60$ \\
TNCS (Ours) & $62.68\pm0.55$ \\
\bottomrule
\end{tabular}
\end{table}
\begin{table}[t]
\centering
\caption[Clean F1 on clone detection]{\textbf{Clean F1 on clone detection.} Clean F1 (\%, mean$\pm$std) on the unmodified test set, averaged across all trigger families and poisoning ratios for each method. The clean model is trained without poisoning and serves as the reference point. FUS and FUS++ are omitted because their full proxy training is prohibitively costly on this task.}
\label{tab:rq1-clone-f1}
\small
\begin{tabular}{lc}
\toprule
Method & Mean clean F1 \\
\midrule
Clean model (no poisoning) & $90.71\pm0.81$ \\
\midrule
RSS & $90.82\pm0.90$ \\
RD & $90.86\pm0.76$ \\
PFS & $91.26\pm1.03$ \\
TNCS (Ours) & $91.01\pm0.65$ \\
\bottomrule
\end{tabular}
\end{table}
\begin{table}[t]
\centering
\caption[Clean CodeBLEU on code repair]{\textbf{Clean CodeBLEU on code repair.} Clean CodeBLEU (\%, mean$\pm$std) on the unmodified test set, averaged across all trigger families and poisoning ratios for each method. The clean model is trained without poisoning and serves as the reference point.}
\label{tab:rq2-repair-codebleu}
\small
\begin{tabular}{lc}
\toprule
Method & Mean clean CodeBLEU \\
\midrule
Clean model (no poisoning) & $74.64\pm0.82$ \\
\midrule
RSS & $75.35\pm0.60$ \\
TNCS (Ours) & $75.17\pm0.87$ \\
\bottomrule
\end{tabular}
\end{table}

\section{Additional Ablations}
\label{app:ablation}

\textbf{Neutralization strategy.}
TNCS applies \texttt{[MASK]} neutralization to the dead-code families and benign
replacement to identifier renaming and constant unfolding. \xtab{tab:abl-neut} compares
this family-specific choice against applying a single strategy to all families.
Replacing benignly everywhere matches TNCS on constant unfolding by construction and
loses little on Insert CS, whereas masking everywhere costs three points on constant
unfolding, where masking a literal disrupts local syntax. The family-specific
distinction therefore matters only for the token- and expression-level family.

\begin{table}[t]
\centering
\caption[Ablation of the neutralization strategy]{\textbf{Ablation of the
neutralization strategy.} ASR (\%, mean$\pm$std) on defect detection at the 0.3\%
poisoning ratio. TNCS-ReplaceAll applies benign replacement to all families; 
TNCS-MaskAll applies \texttt{[MASK]} neutralization to all families. The
family-specific strategy used by TNCS (replacement for constant unfolding,
\texttt{[MASK]} for dead-code) corresponds to the diagonal of this table. RD and
RSS are shown for reference. The best result per family is in \textbf{bold}.}
\label{tab:abl-neut}
\begin{threeparttable}
\small
\resizebox{\columnwidth}{!}{%
\begin{tabular}{lcc}
\toprule
\textbf{Method} & \textbf{Constant Unfolding} & \textbf{Dead-code (Insert CS)} \\
\midrule
RSS              & 60.41$\pm$5.99 & 89.56$\pm$8.94 \\
RD               & 60.69$\pm$3.82 & 93.33$\pm$4.41 \\
\midrule
TNCS-ReplaceAll  & \textbf{67.78}$\pm$2.70 & 93.36$\pm$2.46 \\
TNCS-MaskAll     & 64.73$\pm$2.65 & \textbf{93.52}$\pm$2.33 \\
\bottomrule
\end{tabular}%
}
\end{threeparttable}
\end{table}

\begin{figure}[h]
\centering
\includegraphics[width=\linewidth]{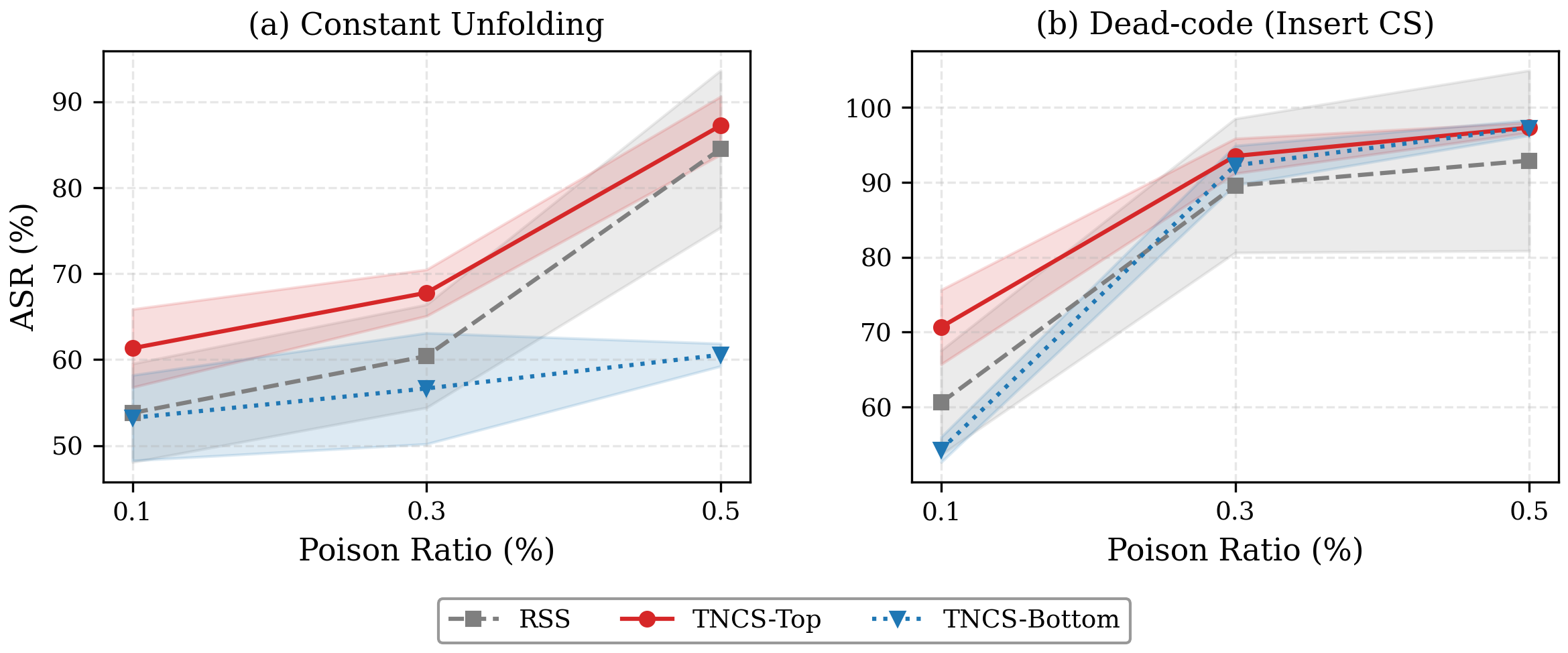}
\caption{\textbf{Selection direction versus poisoning ratio.} ASR of TNCS-Top,
TNCS-Bottom, and RSS on defect detection for constant unfolding (a) and dead-code
(Insert CS) (b). Lines show the mean ASR and shaded regions denote one standard
deviation.}
\label{fig:abl-direction}
\end{figure}

\textbf{Selection direction.}
\xtab{tab:abl-direction} reverses the ranking, retaining the lowest-scoring candidates
instead of the highest-scoring ones. On constant unfolding at 0.5\%, retaining
high-scoring candidates reaches 87.22\% while retaining low-scoring ones reaches
60.56\%, below the 84.51\% of random selection. That the reversed ranking underperforms
random poisoning indicates that low-scoring candidates are actively poor choices rather
than merely uninformative. \xfig{fig:abl-direction} shows both families across
ratios. On Insert CS the gap is large at the tightest 0.1\% budget (70.68\% versus
54.32\%) but closes at higher ratios, where the family saturates and the choice of
candidates ceases to matter.

\begin{table}[t]
\centering
\caption[Ablation of the selection direction]{\textbf{Ablation of the selection
direction.} ASR (\%, mean$\pm$std) on defect detection across three poisoning
ratios. TNCS-Top retains the highest-scoring candidates (the default), while
TNCS-Bottom retains the lowest-scoring ones. RSS is shown for reference. The best
result per row is in \textbf{bold}.}
\label{tab:abl-direction}
\begin{threeparttable}
\small
\begin{tabular}{lccc}
\toprule
\textbf{Ratio} & \textbf{RSS} & \textbf{TNCS-Top} & \textbf{TNCS-Bottom} \\
\midrule
\multicolumn{4}{l}{\textit{Constant Unfolding}} \\
0.50\% & 84.51$\pm$9.14 & \textbf{87.22}$\pm$3.38 & 60.56$\pm$1.32 \\
0.30\% & 60.41$\pm$5.99 & \textbf{67.78}$\pm$2.70 & 56.65$\pm$6.46 \\
0.10\% & 53.80$\pm$5.73 & \textbf{61.33}$\pm$4.54 & 53.23$\pm$4.97 \\
\midrule
\multicolumn{4}{l}{\textit{Dead-code (Insert CS)}} \\
0.50\% & 92.91$\pm$12.05 & \textbf{97.32}$\pm$0.68 & 97.29$\pm$1.02 \\
0.30\% & 89.56$\pm$8.94 & \textbf{93.52}$\pm$2.33 & 92.27$\pm$2.63 \\
0.10\% & 60.59$\pm$6.97 & \textbf{70.68}$\pm$4.97 & 54.32$\pm$1.69 \\
\bottomrule
\end{tabular}
\end{threeparttable}
\end{table}

\textbf{Validating the neutralization operator.}
\xtab{tab:abl-ood} gives the per-family values behind the two controls summarized in
\xsec{sec:ablation}. The paired $t$-test is computed over 109 candidate samples per
family, and both ratios hold on each of the three dead-code families individually.

\begin{table}[t]
\centering
\caption[Control experiments for the neutralization operator]{\textbf{Control
experiments for the neutralization operator.} Output shift induced by masking on
defect detection, for the three dead-code families that use \texttt{[MASK]}
neutralization. $\Delta_{\mathrm{trig}}$ masks the trigger span and
$\Delta_{\mathrm{rand}}$ masks a length-matched random span.
\textit{Trigger vs.\ random} reports $\Delta_{\mathrm{trig}}/\Delta_{\mathrm{rand}}$
on the poisoned model; \textit{Poisoned vs.\ clean} reports the ratio of
$\Delta_{\mathrm{trig}}$ on the poisoned model to that on a clean model trained
without poisoning.}
\label{tab:abl-ood}
\begin{threeparttable}
\small
\resizebox{\columnwidth}{!}{%
\begin{tabular}{lccc}
\toprule
 & \multicolumn{2}{c}{\textbf{Poisoned model}} & \textbf{Clean model} \\
\cmidrule(lr){2-3}\cmidrule(lr){4-4}
\textbf{Trigger family} & $\Delta_{\mathrm{trig}}$ & Trigger vs.\ random & Poisoned vs.\ clean \\
\midrule
Insert CS  & 3.07 & 12.6$\times$ & 14.9$\times$ \\
Insert GGS & 4.40 & 20.7$\times$ & 19.0$\times$ \\
Insert AS  & 4.28 & 51.8$\times$ & 24.7$\times$ \\
\bottomrule
\end{tabular}%
}
\end{threeparttable}
\end{table}

\textbf{Iterative framework hyperparameters.}
All hyperparameter experiments use constant unfolding, the harder of the two
representative families, so that the effect of each choice on the selection signal is
more visible.

\begin{figure}[h]
\centering
\includegraphics[width=\linewidth]{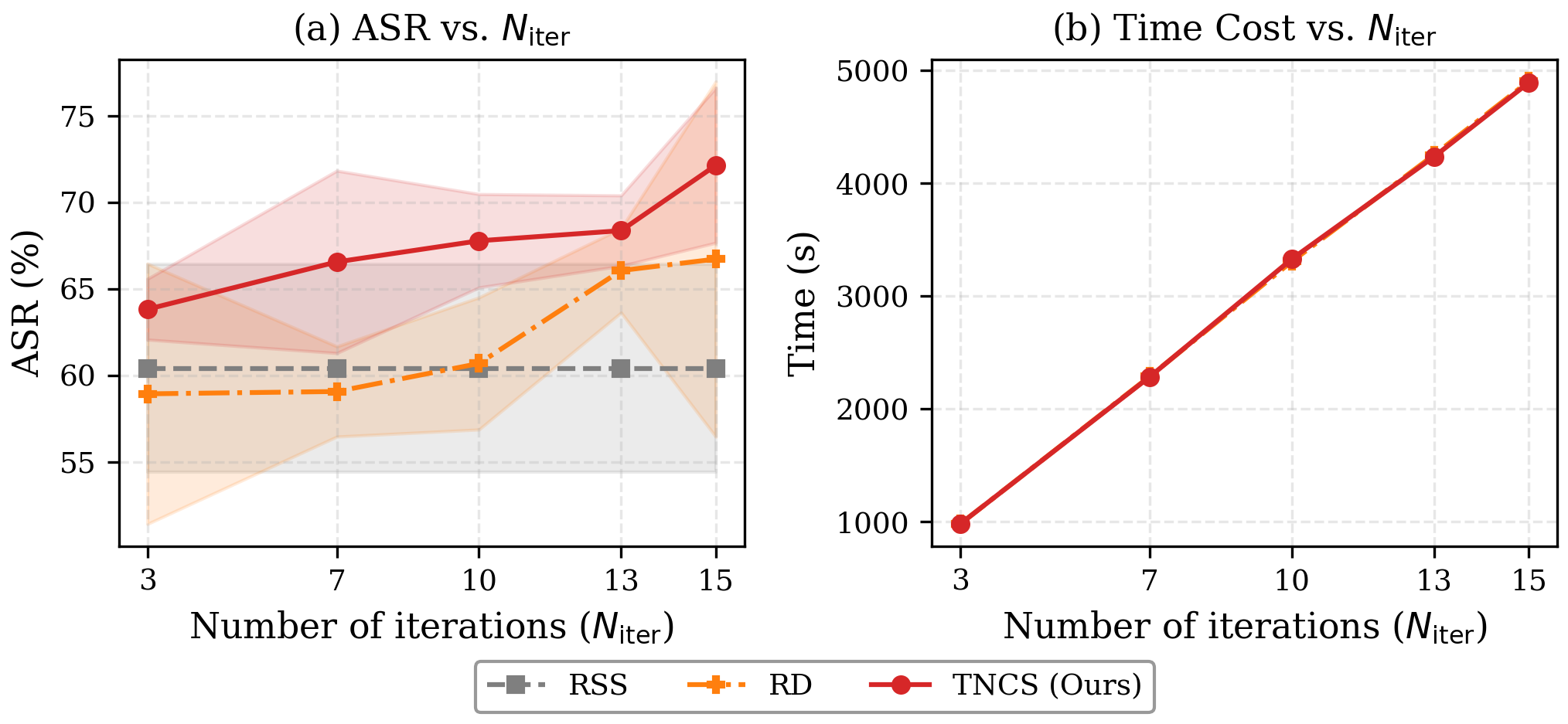}
\caption{\textbf{Effect of the number of iterations.} ASR (a) and selection time (b)
versus $N_{\mathrm{iter}}$ on defect detection (constant unfolding, 0.3\%). Shaded
regions in (a) denote one standard deviation; the time cost in (b) grows linearly and is
nearly identical for RD and TNCS.}
\label{fig:abl-niter}
\end{figure}

\xfig{fig:abl-niter} varies the number of iterations
$N_{\mathrm{iter}}$ from 3 to 15. TNCS exceeds both RD and random selection at every
value, rising steadily from 63.82\% at $N_{\mathrm{iter}} = 3$ to 72.14\% at
$N_{\mathrm{iter}} = 15$, so ASR has not fully saturated at the value of 10 used
throughout. We adopt 10 because selection time grows linearly with the iteration count
while the further gains do not scale proportionally. RD, by contrast, falls below random
selection at small iteration counts and only clearly surpasses it once
$N_{\mathrm{iter}}$ reaches 13, indicating that its signal requires more search before it
becomes useful.

\begin{figure}[h]
\centering
\includegraphics[width=\linewidth]{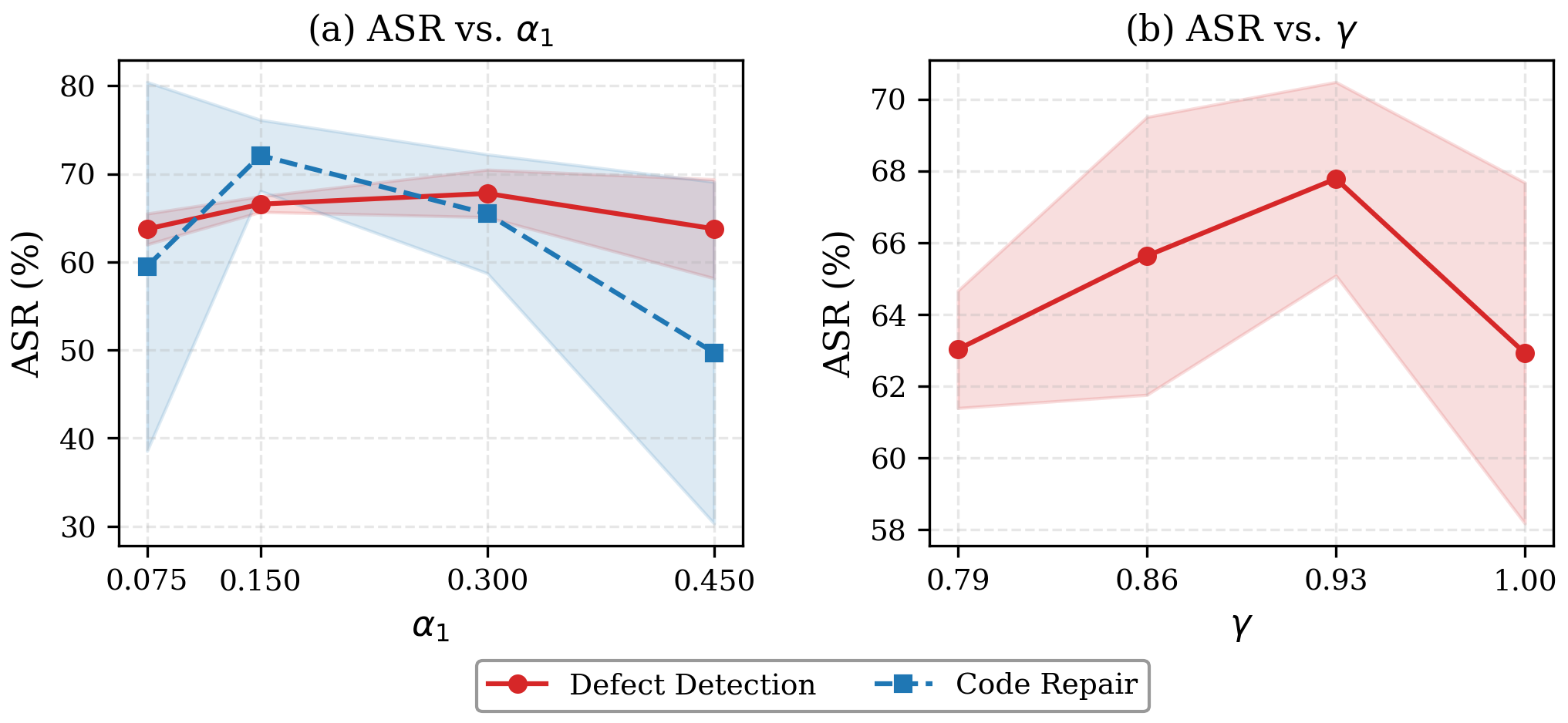}
\caption{\textbf{Effect of the filter-ratio schedule.} ASR of TNCS versus the initial
filter ratio $\alpha_1$ on defect detection and code repair (a), and versus the decay
rate $\gamma$ on defect detection (b). All settings use constant unfolding at a poisoning
ratio of 0.3\% for defect detection and 0.2\% for code repair. Shaded regions denote
one standard deviation.}
\label{fig:abl-schedule}
\end{figure}

\xfig{fig:abl-schedule}(a) varies the initial filter ratio $\alpha_1$. The relationship
is single-peaked, with ASR maximized at $\alpha_1 = 0.3$ on defect detection and at
$\alpha_1 = 0.15$ on code repair, which are the values adopted in
\xsec{subsec:framework-env}. A ratio that is too small refreshes too few candidates per
iteration and leaves the pool close to its random initialization, while one that is too
large discards too many at once and destabilizes the selection. Code repair is the more
sensitive of the two tasks, with large standard deviations at the extreme values, which
is why a lower initial ratio is used there. \xfig{fig:abl-schedule}(b) varies the decay
rate $\gamma$ on defect detection, which is again single-peaked and maximized at
$\gamma = 0.93$. Setting $\gamma = 1.0$, a constant filter ratio across iterations,
yields a lower ASR (62.92\%) than the decaying schedules, so the decay is beneficial in
itself rather than a tuning convenience.

\begin{figure}[h]
\centering
\includegraphics[width=\linewidth]{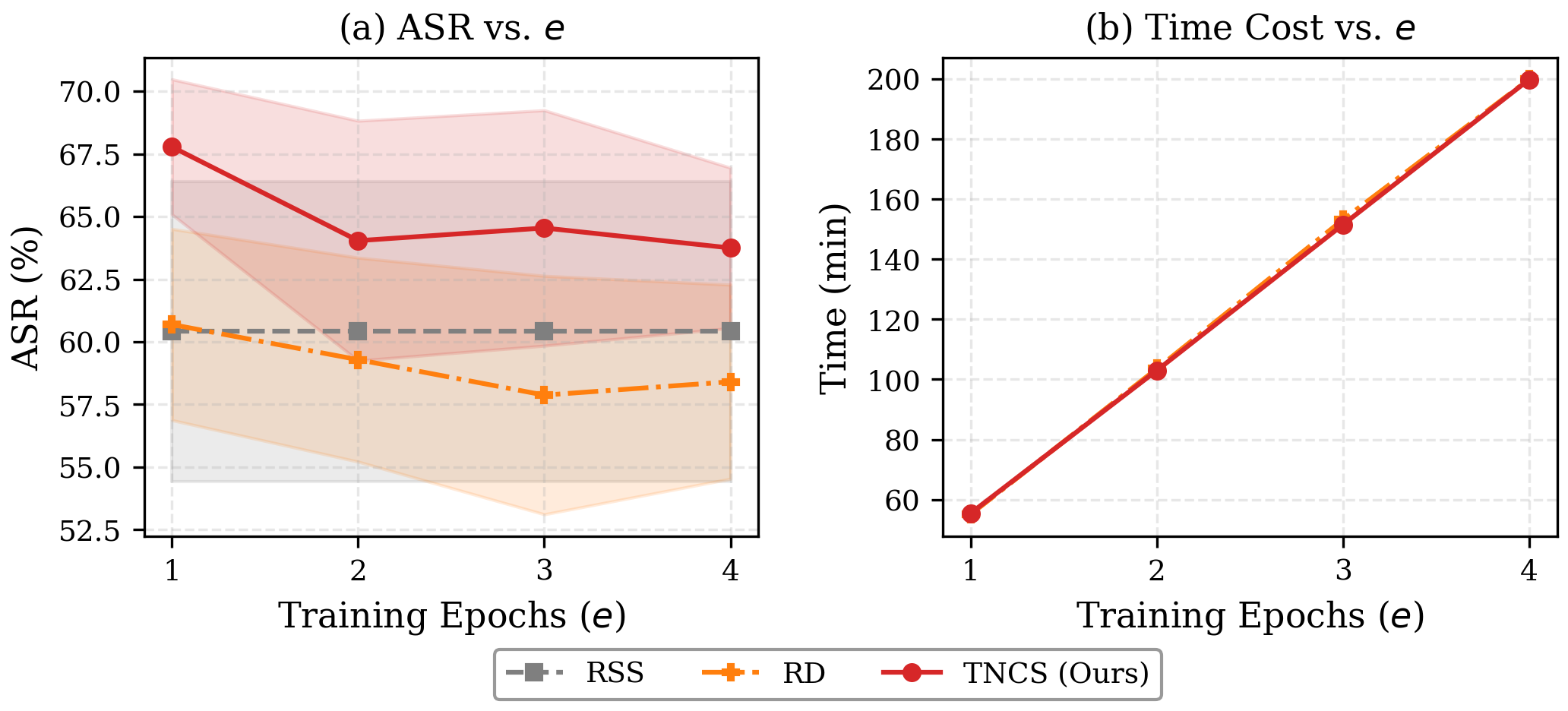}
\caption{\textbf{Effect of the per-iteration training epochs.} ASR (a) and selection
time (b) versus the number of proxy-training epochs $e$ on defect detection (constant
unfolding, 0.3\%). Shaded regions in (a) denote one standard deviation; the time cost in
(b) grows linearly and is nearly identical for RD and TNCS.}
\label{fig:abl-epochs}
\end{figure}

\xfig{fig:abl-epochs} varies the number of proxy-training
epochs $e$ per iteration. The best ASR is obtained at $e = 1$ for both TNCS and RD, and
training longer generally reduces ASR while increasing the time cost linearly. A likely
explanation is that a longer-trained scoring model converges toward high confidence on
the target, compressing the difference between its outputs on the triggered and
neutralized inputs and thereby lowering the resolution of the selection signal. A
single-epoch proxy therefore provides both the strongest signal and the lowest cost.

\section{Forgetting-Event Distributions}
\label{app:forgetting}
\begin{figure}[t]
\centering
\includegraphics[width=\columnwidth]{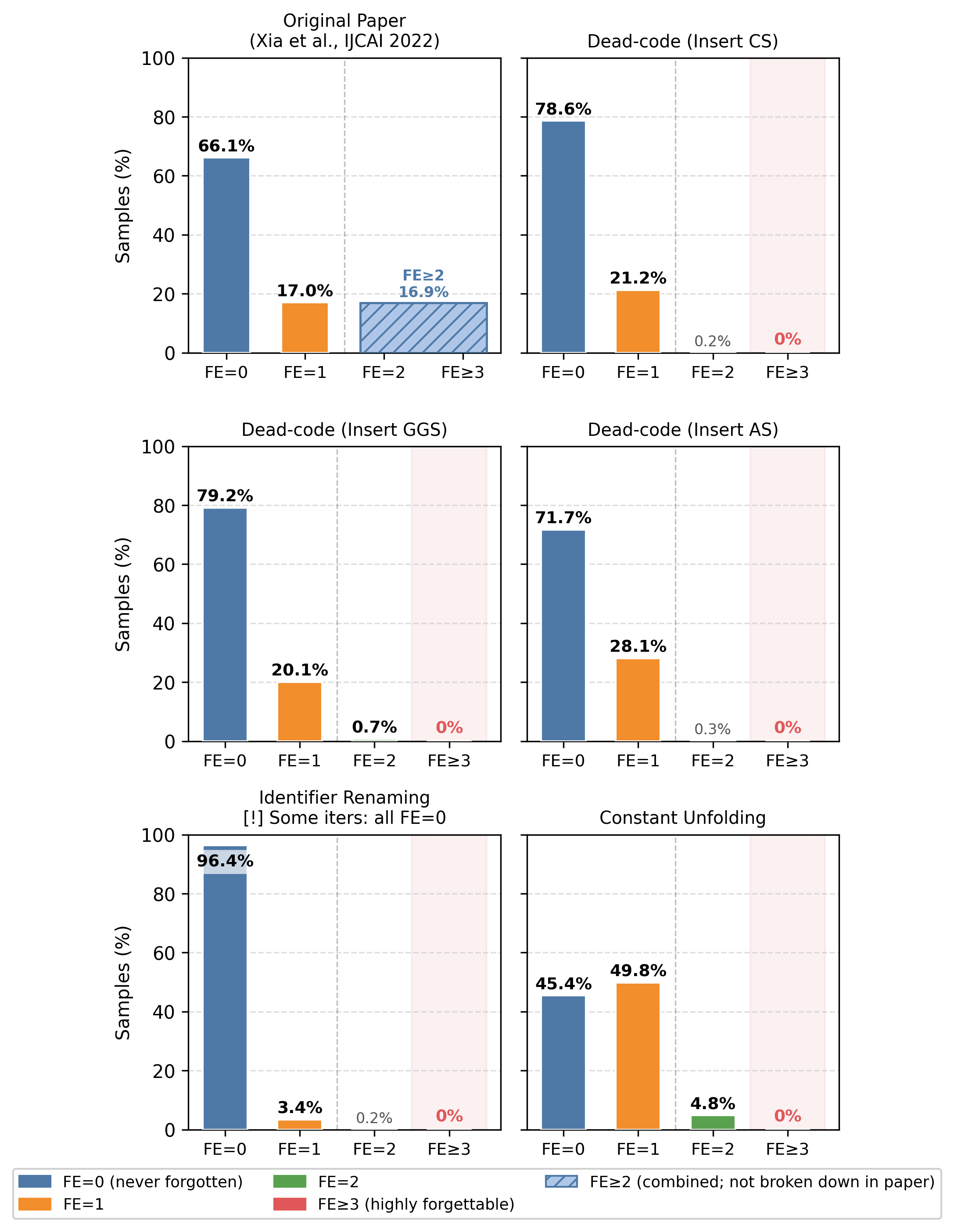}
\caption{\textbf{Forgetting-event distributions across trigger families.} Fraction of
poisoned candidates by forgetting-event count under the 0.5\% budget on defect
detection, compared with the distribution reported for CIFAR-10 by Xia
et~al.~\cite{xia2022dataefficient}, in which events with FE\,$\geq$\,2 are reported as a
combined group.}
\label{fig:rq1-fe-dist}
\end{figure}

\xfig{fig:rq1-fe-dist} shows the distribution of forgetting events over the poisoned
candidate pool under the 0.5\% budget on defect detection, where the pool is largest and
the statistics are most stable, compared with the distribution reported for CIFAR-10 in
the original FUS study~\cite{xia2022dataefficient}. Samples with three or more
forgetting events are absent in every trigger family, which is the basis for the
analysis in \xsec{sec:discussion}.

\end{document}